\documentclass[conference,compsoc]{IEEEtran}
\ifCLASSOPTIONcompsoc
  \usepackage{cite}
\else
  \usepackage{cite}
\fi
\ifCLASSINFOpdf
\else
\fi
\usepackage{lipsum}
\usepackage{xcolor}
\usepackage[table]{xcolor}
\usepackage{enumitem}
\usepackage{comment}
\usepackage{xspace}
\usepackage{amsmath}
\usepackage{graphicx}
\usepackage{listings}
\usepackage[hyphens]{url}
\usepackage{hyperref}
\hypersetup{                    
  colorlinks,
  urlcolor={blue!70!black}
}
\hypersetup{breaklinks=true}
\usepackage[noabbrev,capitalize]{cleveref}
\usepackage{makecell}
\usepackage{cellspace}
\usepackage{multirow}
\usepackage{amssymb}
\usepackage{amsbsy}
\usepackage{tabularx}
\usepackage[export]{adjustbox}

\usepackage{subcaption}
\usepackage{booktabs}
\usepackage{seqsplit}
\usepackage{tcolorbox}
\usepackage{tikz}
\usetikzlibrary{arrows.meta, positioning, calc}
\usepackage{tikzpagenodes}
\usepackage[available,functional,reproduced]{ieeebadges}
\newcommand{\TODO}[1]{{\footnotesize\color{red}[TODO: #1]}}

\newtcolorbox{observation}[1][]{
    colback=black!10, colframe=black!85!black, coltitle=black,
    left=1mm, right=1mm, top=1mm, bottom=1mm, boxrule=0.5mm,
    #1
}

\newcommand{\papername}{GPUBreach\xspace}
\definecolor{codegreen}{rgb}{0,0.6,0}
\definecolor{codegray}{rgb}{0.5,0.5,0.5}
\definecolor{codepurple}{rgb}{0.58,0,0.82}
\definecolor{backcolour}{RGB}{245,245,245}
\definecolor{keywordcolour}{RGB}{173, 16, 16}
\lstdefinestyle{mystyle}{
    language=bash,
    backgroundcolor=\color{backcolour},   
    commentstyle=\color{codegreen},
    keywordstyle=\color{keywordcolour},
    numberstyle=\tiny\color{codegray},
    stringstyle=\color{codepurple},
    basicstyle=\ttfamily\bfseries\footnotesize,
    breakatwhitespace=false,         
    breaklines=true,                 
    captionpos=b,                    
    keepspaces=true,                 
    numbers=left,                    
    numbersep=5pt,                  
    showspaces=false,                
    showstringspaces=false,
    showtabs=false,                  
    tabsize=2,
    frame=single,
    morekeywords={size_t},
    rulecolor=\color{black},
}

\tikzset{
  revision box/.style args={#1/#2}{
    draw=#1,
    text=#1,
    font=\bfseries\color{#1},
    rounded corners,
    minimum height=0.5cm,
    text width=1.3cm,
    align=center,
    thick,
    fill=#2!20,
    inner sep=2pt
  }
}

\newcommand{\REVISIONTEXTCOLOR}{black}
\newcommand{\REVISIONBOXCOLOR}{black}
\newcommand{\REVISIONFILLCOLOR}{red}
\newcommand{\REVISION}[1]{{\color{\REVISIONTEXTCOLOR}#1}}

\NewDocumentCommand{\revisionbox}{O{} m m}{%
\begin{tikzpicture}[overlay, remember picture, baseline]
\coordinate (revpos) at ($(0,0)$);
\node[revision box=\REVISIONBOXCOLOR/\REVISIONFILLCOLOR,](revbox) 
    at ($(current page text area.north east |- revpos) + #1$){#3};
\draw[\REVISIONFILLCOLOR, line width=1.5pt]
        (revbox.south) -- ++(0, -8pt);
    \draw[\REVISIONFILLCOLOR, line width=1.5pt]
        (revbox.south) ++(0, -8pt) -- ++(-\columnwidth-#2, 0);
\end{tikzpicture}
}

\newcommand{\leftx}{-18.7cm}
\newcommand{\lefty}{0.45cm}
\newcommand{\rightx}{1cm}
\newcommand{\righty}{0.45cm}
\newcommand{\rightlengthofs}{\rightx}
\newcommand{\leftlengthofs}{\leftx - 0.6cm}

\newif\ifrevisions
\revisionsfalse 

\ifrevisions
  \newcommand{\rightrevisionbox}[1]{\revisionbox[(\rightx, \righty)]{\rightlengthofs}{#1}}
  \newcommand{\leftrevisionbox}[1]{\revisionbox[(\leftx, \lefty)]{\leftlengthofs}{#1}}
\else
  \newcommand{\rightrevisionbox}[1]{}
  \newcommand{\leftrevisionbox}[1]{}
\fi

\begin{document}
%
\title{GPUBreach: Privilege Escalation Attacks on GPUs using Rowhammer}

\author{\IEEEauthorblockN{Chris S. Lin, Yuqin Yan, Guozhen Ding, Joyce Qu, Joseph Zhu, David Lie, Gururaj Saileshwar}
\IEEEauthorblockA{
 University of Toronto}
}


%


\maketitle

\begin{abstract}
NVIDIA GPUs with GDDR memories have been shown susceptible to Rowhammer-based bit-flips, similar to CPUs. However, Rowhammer exploits on GPUs have been limited to injecting untargeted bit-flips in victim data like weights of machine learning models, to degrade model accuracy, unlike CPU exploits shown capable of privilege escalation. 

In this paper, we demonstrate that GPU Rowhammer exploits can be as potent as CPU Rowhammer attacks. By exploiting the GPU page table management to identify when and where new page tables are allocated, we enable an unprivileged user CUDA kernel of one process to use RowHammer bit-flips to gain access to the GPU memory of other processes or co-tenants via targeted tampering of such page-tables resident on the GPU memory. Using this newly found primitive, we demonstrate the first GPU-side privilege escalation attacks, leaking secret data such as cryptographic keys from cuPQC libraries, and even tampering with the model's GPU assembly code to degrade models more stealthily than previous attacks. We further demonstrate that GPU-side privilege escalation can lead to CPU-side privilege escalation, defeating the protections provided by the IOMMU, enabling a malicious user-level program with GPU access to gain root shell and system-wide control, even in a non-multi-tenant setting. 

\end{abstract}


%
\IEEEpeerreviewmaketitle

\section{Introduction}
Modern DRAM is vulnerable to Rowhammer attacks~\cite{Rowhammer2014}, where an attacker can induce bit-flips in memory cells by repeatedly activating neighboring rows. Rowhammer has been extensively studied on CPU DRAM (DDR3--5 and LPDDR4--5) across Intel, AMD, and ARM CPUs~\cite{TRRespass, SMASH, Blacksmith, HalfDouble, Eccploit, ZenHammer, eccfail, GrandPwning, phoenix}, and more recently on GPUs with GPUHammer~\cite{gpuhammer} demonstrating bit-flips on GDDR6 memories.
While CPU based exploits have been extensively studied, and shown capable of weakening cryptographic keys~\cite{crowhammer,PQHammer,FFS}, corrupting \texttt{sudo} to gain supervisor privileges~\cite{AnotherFlip,Eccploit}, or tampering with page tables to obtain kernel-level privileges~\cite{ProjectZeroRowhammer,TRRespass,RowhammerJS,Eccploit,Blacksmith,HalfDouble}, there has been a limited study of Rowhammer exploits on GPUs.
So far, GPU Rowhammer exploits have been only shown to have limited capabilities like tampering with co-resident ML model weights, to degrade model accuracy~\cite{gpuhammer}  or jailbreak large language models~\cite{PrisonBreak}, making it unclear if they can be as potent as CPU-based exploits.

This paper explores whether privilege escalation exploits are feasible on NVIDIA GPUs using Rowhammer.
On NVIDIA GPUs, user code runs as CUDA kernels that allocate memory resident in the GPU. Kernels from different processes (CUDA contexts)
can run in a time-sliced manner
in a single-tenant GPU (e.g., workstation), while 
 kernels from different users can run concurrently, spatially sharing the GPU, via NVIDIA’s Multi-Process Service (MPS) in containerized deployments
 ~\cite{GKE,AlibabaCloudGPU}.
A CUDA kernel with elevated privileges can arbitrarily read or modify another process’s GPU memory (e.g., leak or tamper with cryptographic keys, ML model weights, etc.).
Worse, it may potentially corrupt the CPU-side data of the GPU driver, that executes on the CPU at an elevated privilege, enabling system-wide privilege escalation even on a single-tenant machine.
Thus, GPU privilege escalation can have serious implications in both single and multi-tenant settings.

Privilege escalation exploits on CPUs~\cite{ZenHammer,ProjectZeroRowhammer} use Rowhammer to corrupt privileged binaries (e.g., \textit{sudo}), or to tamper with page tables.
While GPUs do not execute privileged binaries like \textit{sudo}, 
GPU page tables are stored entirely in GPU memory, making them appealing targets for GPU-based Rowhammer. 
Here, the attacker’s  goal is to gain arbitrary read/write access to GPU physical memory by corrupting their own GPU page tables (PT).
Achieving this requires: (1) a page table entry (PTE) to be tampered with a bit-flip, and (2) the corrupted page-frame number (PFN) to point to another page table page. This enables the attacker's CUDA kernel to read and write its own page table, controlling its GPU virtual-to-physical mappings and enabling arbitrary GPU memory access.
However, obtaining this level of control over GPU page tables is nontrivial and faces several challenges ({\bf C1--C3}).

\smallskip
\noindent
\textbf{C1. Where are GPU Page Tables Allocated?}
Limited documentation exists on where NVIDIA GPUs place page tables in device memory. For Rowhammer based PTE tampering to work, a page table page needs to be sandwiched between two user data pages in neighboring rows.
If the GPU allocates page tables in a separate memory region, then the attack cannot succeed.
Thus, we must first determine how NVIDIA GPUs allocate page tables in memory to ensure interspersed allocation of data and page table pages. 

\smallskip
\noindent
\textbf{C2. Populating GPU Memory with PTEs.}
CPU-based attacks often use \texttt{mmap} to map a single physical page to many virtual pages, forcing the memory to fill with page tables. \REVISION{However, the equivalent in GPUs (\texttt{cuMemMap}) fails to achieve the same effect, making this strategy ineffective.}\rightrevisionbox{M2} We thus require new GPU mechanisms to efficiently fill PTEs in memory, while minimizing the data memory requirements.


\smallskip
\noindent
\textbf{C3. Positioning GPU PTEs at Target Locations.}
A successful exploit requires not only flipping a bit in PTEs, 
but also ensuring the corrupted PFN points to another  page-table page with valid PTEs. This requires placing a page-table page precisely at the location that a tampered PFN will reference. How do we reliably steer GPU page-table allocations to these target physical addresses?
In this paper, we develop techniques to overcome these challenges and successfully launch Rowhammer-based privilege escalation exploits on NVIDIA GPUs.
Similar to prior work~\cite{gpuhammer}, we demonstrate these techniques on an NVIDIA RTX A6000 GPU with GDDR6 memory, that is widely available in workstations and cloud servers, and shown to be vulnerable to Rowhammer bit-flips.

\smallskip
\noindent
\textbf{Uncovering Memory Locations of GPU Page Tables.}
Building on prior reverse engineering of GPU TLBs~\cite{tunnels}, we patch the open-source NVIDIA GPU driver to dump the memory locations of GPU page tables and analyze its allocation strategy. We find that the driver places all page tables in contiguous 2MB regions, which we call \textit{page table regions} (PT regions). 
However, the initial PT region is far from user data. Only after this initial region fills, will a new region be allocated from the same physical memory pool as user pages, enabling interspersed allocations required for a Rowhammer exploit.
Thus, we need primitives to efficiently fill PT regions with PTEs, with minimum memory usage.


\smallskip
\noindent
\textbf{Efficient Allocation of GPU Page Table Pages.}
To reliably induce bit-flips in PTEs, 
we need the ability to (1) allocate new PT regions at chosen locations, by filling PTEs in existing PT regions, and (2) densely pack PT regions with valid PTEs for effective Rowhammer targeting.
While GPUs support multiple data page sizes (2~MB, 64~KB, and 4~KB),  
the NVIDIA driver automatically promotes allocations to 2~MB pages~\cite{tunnels}, without providing the user the capability to request specific page frame sizes. 
Consequently, allocating even a single new 2 MB PT region naively consumes roughly 256 GB of memory, far beyond GPU capacity.
To address this, we develop new allocation primitives using Unified Virtual Memory (UVM) that coerce the allocator into producing desirable 64KB and 4KB page frames. Using these primitives, we can allocate new PT regions while consuming only 1 GB of memory and densify existing PT regions for effective Rowhammer targeting.

\smallskip
\noindent
\textbf{Side-Channel for Detecting Page Table Allocations.} To position PT pages at desired locations, 
the attacker must detect \emph{when} a new PT region is allocated.
To that end, we uncover a \textit{new} timing side-channel from UVM allocations that 
trigger evictions to CPU memory, when the GPU is near capacity.
By keeping GPU memory just below full and issuing UVM allocations, the attacker can observe which allocations cause an eviction due to the creation of an extra PT region, thus detecting new PT region allocations.
By controlling which memory region is freed, 
the attacker can guide page-table placement to the desired physical address.

\smallskip
\noindent
\textbf{Exploitation.}
With these techniques, we not only massage a page table page into a location with a known bit-flip, but also a second page table page at the address referenced by the corrupted PFN. Thus, the attacker gains control of its own page table, and gets arbitrary read/write privilege on the GPU memory. We use this capability to demonstrate three real-world exploits on a GPU that is temporally shared by kernels of different user processes.

First, we demonstrate that NVIDIA's cuPQC, a post-quantum cryptography (PQC) library using GPU acceleration, that stores its keys in GPU DRAM during operations such as key-exchange, can have its secret keys leaked by an attacker that runs in a time-sliced manner with a victim. 

Second, we demonstrate code tampering in PyTorch ML model inference. 
By tampering with a single branch in the SASS (GPU assembly) of NVIDIA's cuBLAS library in GPU memory, an attacker can universally degrade model accuracy from up to 80\% down to 0\%, more stealthily than prior works that tamper with model weights~\cite{gpuhammer,TBD,Deephammer,bitflipattack}.
We also show leakage of LLM weights from GPU memory.


Third, we leverage the compromised GPU to issue malicious DMA accesses to the CPU and tamper with the privileged GPU driver memory on the CPU.
By corrupting implicitly trusted, GPU-writable data structures in the privileged driver,
we induce out-of-bounds (OOB) accesses and arbitrary writes within the Linux kernel, even with standard defenses such as IOMMU enabled,~\cite{nvidia_iommu,amd_iommu,msft_iommu} leading to system-wide privilege escalation and a root shell for the attacker.
These results show that GPU Rowhammer attacks threaten not just multi-tenant settings,
but all GPU users.

\smallskip
\noindent Overall, this paper makes the following contributions:
\begin{enumerate}[label={\arabic*)},itemsep=0pt]
  \item We demonstrate the first privilege escalation exploit on NVIDIA GPUs via page table tampering using Rowhammer, granting arbitrary GPU R/W privileges.
 \item We uncover unknown GPU page table allocation behaviors and new side-channels that make page table massaging and tampering practical on NVIDIA GPUs.
 \item Leveraging arbitrary R/W privileges on GPU memory, we demonstrate a wide range of exploits from  stealing or tampering sensitive data (ML models, cryptographic keys) to system-wide privilege escalation.
\end{enumerate}

\smallskip
\noindent
\textbf{Responsible Disclosure.}
We responsibly disclosed our exploit to NVIDIA on 11th November, 2025, and subsequently to Google, AWS, and Microsoft.
\REVISION{Google awarded us a bug bounty award of USD 600. 
NVIDIA responded that they may update their Rowhammer security notice~\cite{NVIDIARowhammerNotice}.
Our code is open-sourced at \url{https://github.com/sith-lab/gpubreach}.}
\section{Background}
\subsection{GPU  Privilege Levels}

Modern GPUs do not follow the traditional privilege-ring model of CPUs. Instead, their management is divided between proprietary user-level code, the kernel-mode driver on the CPU, and privileged on-device microcontrollers.
Consequently, we define a party as \textit{privileged} on the GPU if it possesses capabilities equivalent to the kernel-mode driver, such as  arbitrary access to GPU memory or the ability to issue  control commands.
Next, we describe the role of the user-level runtime, the kernel-mode driver, and the on-device privileged hardware in NVIDIA GPUs.


\smallskip\noindent
\textbf{User-Level Components.} 
The application code, written in CUDA or other GPU programming frameworks, executes with the lowest privilege level on the GPU. Such code runs on the GPU’s CUDA cores under isolation enforced by the driver and hardware. On the host side, user applications interact with the GPU exclusively through runtime and driver APIs, which mediate all resource allocation, kernel launches, and memory operations. 
On the device side, CUDA kernels are confined to GPU virtual address spaces set up by the driver, preventing them from accessing data of other processes or privileged data governing GPU functionality.

\smallskip
\noindent\textbf{Kernel Driver.} The GPU kernel-mode driver operates in the host’s privileged (kernel) mode and acts as the primary mediator between the user and the GPU hardware. It manages GPU memory allocation, context switching, command submission, and I/O data transfers. Importantly, the driver controls the virtual memory management for the GPU for a given process: it allocates GPU page tables, establishes mappings between GPU virtual addresses and physical memory, and enforces process isolation on the GPU-side.


\smallskip
\noindent
{\color{black}{\textbf{Privileged Hardware Components.} Modern GPUs include privileged hardware components responsible for GPU management. The GPU System Processor (GSP) is an embedded co-processor on the device running trusted firmware for GPU initialization, resource management, and configuration. The GPU Memory Management Unit (GMMU) enforces virtual address translation, memory protection policies, and address space isolation between clients. These components communicate with the privileged driver via DMA to manage the execution environment of the GPU.}}

\smallskip
In this paper, we investigate whether a malicious CUDA code executing at the user-level on the GPU, can achieve privilege-escalation on the device-side, and access arbitrary VRAM of other processes (equivalent to the kernel mode driver), and worse, if it can escalate privileges to the OS-kernel level on the host side, to get system-wide control.

\subsection{GPU Virtual Memory \& Page Tables}

\begin{figure}[bth]
\centering
\includegraphics[width=3.3in,height=\paperheight,keepaspectratio]{"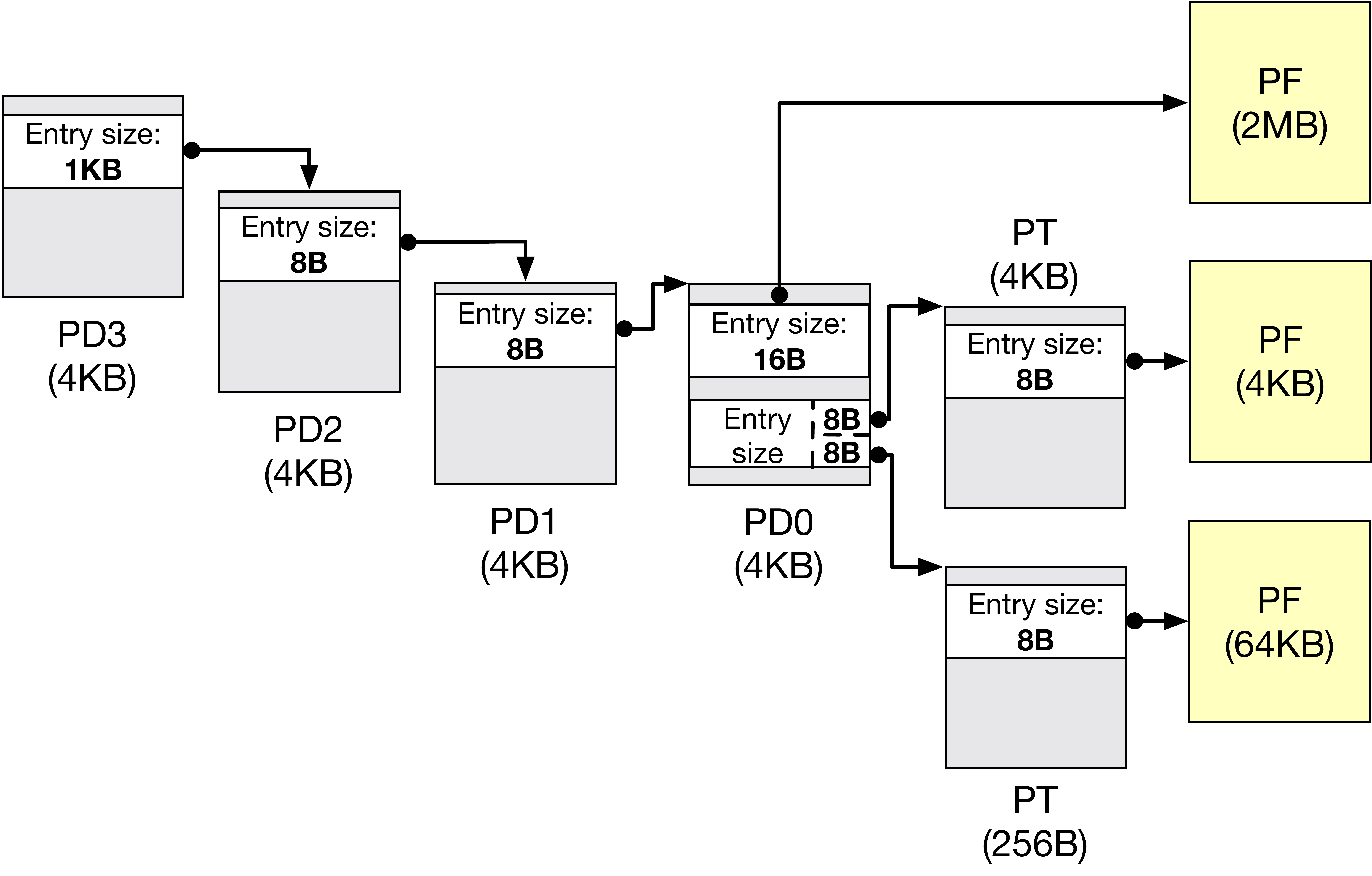"}
\caption{NVIDIA GPU Page Table. It has 4 or 5 levels, including PD3 to PD0 for all data pages, and a PT level only for 4KB and 64KB data pages.}
\label{fig:page_table}
\end{figure}

\noindent
\textbf{Virtual Memory.} On NVIDIA GPUs, each process running CUDA kernels has a distinct GPU context with a unique GPU virtual address (VA) space.
At runtime, on accessing a GPU VA, 
the GMMU performs a virtual-to-physical translation using the GPU page table. 
Each GPU context has distinct page tables stored in the GPU memory. 
As shown in \cref{fig:page_table}, the NVIDIA GPU page table \REVISION{has a multi-level design}, with page directories (PD3-PD0), page tables (PT), and page table entries (PTEs) like the CPU page table. 

\vspace{0.1in}
\noindent
\textbf{Page Tables.} Modern GPUs, starting from the Turing generation (including the Ampere GPU we study), support multiple page sizes, including 2MB, 64KB, and 4KB page frames~\cite{tunnels,nayak2021mis}.
For 2MB page frames, the translation is stored directly in 16B entries at the PD0 level. \REVISION{For 4KB and 64KB page frames, the 16B PD0 entry is divided into two 8B halves, each pointing to the last-level page table (PT). The PT sizes are 4KB and 256B for 4KB and 64KB page frames, respectively.}\rightrevisionbox{M3}
CUDA memory allocation APIs do not provide a way to request specific page sizes. The driver uses 2MB pages for \texttt{cudaMalloc}'ed pages and while it initially allocates smaller pages for unified virtual memory (UVM), these are also aggressively coalesced to 2MB pages~\cite{tunnels}.

\begin{figure}[ht]
\centering
\includegraphics[width=3.3in,height=\paperheight,keepaspectratio]{"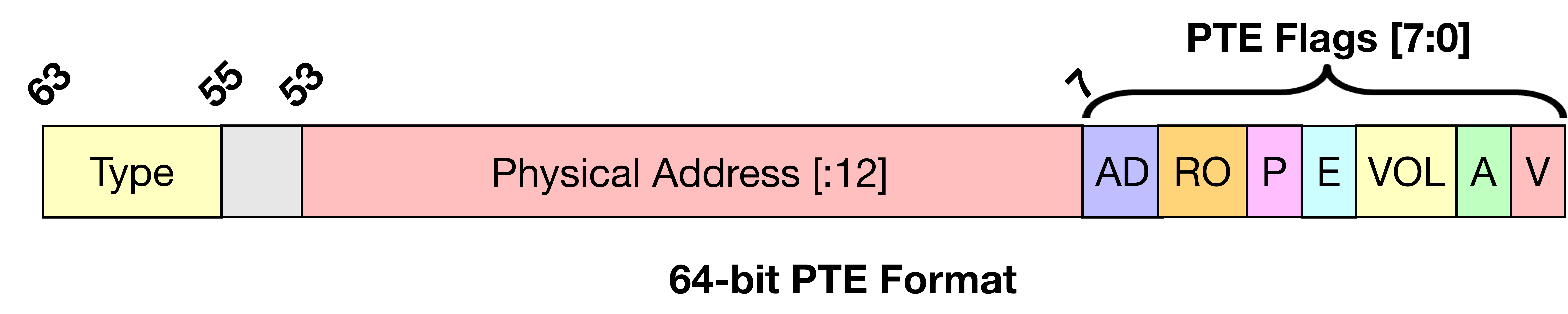"}
\caption{Format of GPU Page Table Entries. The bottom 8 bits are metadata flags, and the next 46 bits contain the page-frame-number.}
\label{fig:pte}
\end{figure}

\noindent
\textbf{Page Table Entries (PTEs).} 
 \cref{fig:pte} shows the format of a PTE in GPU Page Tables (PT) \cite{Pascal_MMU_Format}. 
The 8th to 53rd bits store the page frame number, i.e., the physical address without the bottom 12 bits, as the minimum page size is 4KB. This gives us 58 bits of addressable physical memory. The PTEs also contain metadata flags; some notable ones are the RO (read-only bit), A (2-bit aperture, which indicates the memory domain for the page, 00 and 01 correspond to VRAM, and 10 and 11 correspond to coherent and non-coherent system memory), and P (privilege bit).

\begin{figure*}[!htb]
    \centering
\includegraphics[width=7in]{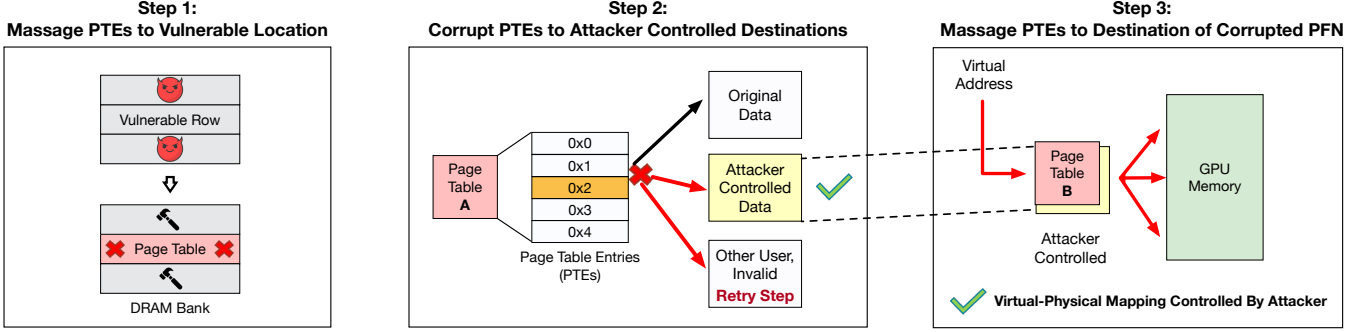}

    \caption{\textbf{Overview of the \papername{} attack}. Steps to tamper GPU page tables with Rowhammer bit-flips to achieve GPU privilege escalation.
    }
    \label{fig:attack_procedure}
\end{figure*}

\subsection{Unified Virtual Memory}\label{subsec:background_uvm}
In modern NVIDIA GPUs, unified virtual memory (UVM) enables a shared virtual address space between CPU and GPU. To enable access to this memory from both sides, the GPU driver transparently migrates physical pages between the GPU and CPU when accessed from the other. 
This avoids the application having to explicitly copying memory from CPU to GPU and vice versa, via \texttt{cudaMemcpy}.

UVM adopts a lazy loading strategy, where data is physically placed in GPU memory only when it is first accessed. When the total UVM allocated memory (via \texttt{cudaMallocManaged}) exceeds the GPU’s physical capacity, the driver avoids out-of-memory errors by evicting inactive GPU-resident pages back to host memory to make room for newly accessed data.
The evictions of GPU resident pages occur in Least-Recently-Used (LRU) order.

\subsection{Rowhammer}
Modern DRAM stores data in cells, with one cell per bit, stored as electrical charge in capacitors arranged in rows and columns. The charge determines the cell's bit value, either 0 or 1. A data access to the DRAM involves activating its corresponding data row, and bringing it into a row buffer. Rowhammer is a read-disturbance phenomenon in which rapid activations of an aggressor row in DRAM causes charge leakage in an adjacent victim row, ultimately inducing bit-flips~\cite{Rowhammer2014}. This vulnerability gets worse as DRAM scales  to smaller technology nodes~\cite{RevisitRowhammer}, and new variants of such data-disturbance vulnerabilities such as Rowpress~\cite{RowPress}, ColumDisturb~\cite{ColumnDisturb}, and Half-Double~\cite{HalfDouble} have continued to surface in recent years. 
Prior works~\cite{TRRespass, SMASH, Blacksmith, HalfDouble, Eccploit, ZenHammer, RevisitRowhammer,Drammer,phoenix, eccfail}, have shown that DDR3-5 and LPDDR4-6 memories, on Intel, AMD, and ARM CPUs are vulnerable to Rowhammer bit-flips, including those with on-die and system-level error correction codes (ECC).

\smallskip
\noindent \textbf{GPU Rowhammer.} 
While Rowhammer attacks have been extensively studied on CPU-based systems, recent research, GPUHammer\cite{gpuhammer}, has shown its feasibility on GPUs. GPUHammer triggered bit-flips on NVIDIA GPUs such as RTX A6000 with GDDR6 memory, by reverse engineering the virtual-physical DRAM row mapping and developing specialized hammering patterns for GPUs that maximize the hammer intensity, while still being synchronized to refreshes~\cite{SMASH} and many-sided~\cite{TRRespass} to bypass in-DRAM mitigations like TRR~\cite{UncoverRowhammer}. Using these bit flips, GPUHammer demonstrates untargeted flips in machine learning model weights~\cite{TBD}, which significantly degrade model accuracy.

Compared to the large number of CPU-based Rowhammer exploits, that demonstrate privilege escalation, by tampering with page tables or instructions in \textit{sudo} binaries~\cite{ProjectZeroRowhammer,TRRespass,RowhammerJS,Eccploit,Blacksmith,HalfDouble,AnotherFlip} or compromising web browsers\cite{GrandPwning, SMASH}, the capability of GPU Rowhammer attacks~\cite{gpuhammer} has been limited to tampering with victim data such as ML model weights to degrade model accuracy.
To that end, this paper studies the potential for GPU privilege escalation attacks, by utilizing Rowhammer bit-flips on GDDR memories.


\section{Overview of \papername{} \& Challenges}

\subsection{Threat Model.}
\textbf{Setting.} We target GPUs with GDDR6 memories, which have been shown vulnerable to Rowhammer in prior works~\cite{gpuhammer}.
Our primary focus is on GPUs natively connected to the host, like in workstations (used by video editors,  gamers, etc.), or used in multi-tenant containerized settings (e.g., GPU shared between containers managed by Kubernetes~\cite{GKE, AlibabaCloudGPU}).
We demonstrate our attacks on the RTX A6000 GPU, for which open-sourced attack code is available~\cite{gpuhammergithub}, although our exploitation methods are generally applicable to any workstation GPU susceptible to bit-flips.\footnote{
NVIDIA's security notice on Rowhammer~\cite{NVIDIARowhammerNotice} suggests that GPUs from multiple generations (Ampere, Ada, Hopper, Blackwell) may be affected. While we focus on the Ampere A6000 GPU, future works may develop attack patterns to defeat in-DRAM defenses on other newer GPUs.}
We do not focus on vGPUs or Multi-Instance GPUs (MIG), since they are primarily used in server-class GPUs with HBM2-3 memories that do not have demonstrated Rowhammer attacks as yet.

We assume Error-Correction-Code (ECC) is disabled on the A6000 GPU. 
This is reflective of real world usage, since we observe two cloud service providers (names redacted to protect customers) providing A6000 and A4000 GPU instances with ECC disabled by default. 
This is because enabling ECC induces up to 10\% slowdown and 6\% memory capacity loss on GPUs with GDDR memories, as shown in prior work~\cite{IMTSullivan,gpuhammer}.
Users in single-tenant settings or those not running ML applications, may keep ECC disabled to avoid these overheads,
as these settings have so far not been shown vulnerable to GPU Rowhammer exploits~\cite{gpuhammer}.

\smallskip
\textbf{Attacker Capabilities.} The attacker is assumed to have the capability to run user-level CUDA code. 
Knowing the target GPU model, the attacker can reverse engineer the virtual address to DRAM row mappings, craft attack patterns that hammer the GPU memory with the highest intensity, and defeat in-DRAM mitigations, as shown in prior work~\cite{gpuhammer}.
The attacker can then identify vulnerable DRAM rows with bit-flips on the victim GPU. 
The attacker's goal is to use these bit-flips to tamper with GPU page tables and escalate privileges on GPU-side to get arbitrary read/write access to GPU memory, to be able to steal or tamper with sensitive data on the GPU, or gain system-wide control. 


\subsection{Overview of the \papername{} Attack}
\papername{} enables privilege escalation attacks on GPUs by tampering with GPU page tables using Rowhammer bit-flips. The crux of the attack is to inject bit-flips in the attacker's GPU PTE, specifically in the Page Frame Number (PFN) field, to redirect the attacker's virtual address to a page table page. As a result, the attacker gets the ability to fully modify their virtual to physical mappings, allowing it arbitrary read/write access to the entire GPU memory. 

The attack consists of 3 key steps, as shown in \cref{fig:attack_procedure}: 

\begin{enumerate}
    \item \textbf{Massage PTEs to a Vulnerable Location.} \papername{} requires \textit{massaging PT regions} to place them precisely at vulnerable memory locations.
    \REVISION{This is because naively using \texttt{cuMemMap} to duplicate virtual pages and fill the memory with PTEs, like prior CPU exploits that use \texttt{mmap}~\cite{ProjectZeroRowhammer},  only allocates up to 4 PT regions ($<$8 MB of PTEs) on the GPU, after which it crashes the program.
    Thus, {\em massaging} PT regions, i.e., precisely placing them in vulnerable memory locations, is needed,}\leftrevisionbox{M2} which requires uncovering the page table allocation algorithm, gaining control of a vulnerable memory page frame, and releasing it exactly when the page-table page is to be allocated, so that PTEs get allocated at the vulnerable memory location.
    \item \textbf{Corrupt PTEs to Attacker Controlled Destinations.} 
    Subsequently, the attacker hammers neighboring rows, to induce bit-flips in the PTE's PFN field. 
    On the first attempt, the corrupted PFN's destination is unlikely to be a PT region, and is rather likely to be an invalid page, or even another user's page.
    Steps 1 and 2 are repeated until the corrupted PFN destination happens to be another data page controlled by the attacker.
    \item \textbf{Massage PTEs to Destination of Corrupted PFN.} Once we identify that the corrupted PFN's destination is an attacker-controlled data page, this data page frame is freed and a PT region is \textit{massaged} to this page frame, enabling the attacker to control PTEs, resulting in GPU privilege escalation. 
\end{enumerate}

A key requirement for this attack is being able to precisely massage page table entries to desired locations, in both Step 1 and 3. This introduces significant challenges.

\subsection{Challenges in Page Table Massaging}
\label{sec:challenges}
\subsubsection*{Challenge 1: Unknown PT Allocation Algorithm} 
The ability to predict \textit{where} page tables are allocated on GPUs is critical for any attempt to manipulate or massage their placement. Unlike CPU operating systems, with well-documented or reverse engineered MMUs, GPU drivers use proprietary and undocumented algorithms to allocate and organize page table pages in device memory. While prior work~\cite{tunnels} demonstrated the capability to dump GPU page tables by instrumenting the GPU driver, key aspects such as their location in memory, allocation granularity, and memory reuse strategies remain unknown. To address this, we empirically study locations of page tables in GPU memory and develop mechanisms to influence their placement, in  \cref{sec:pt_placement}.

\subsubsection*{Challenge 2: Impractical Memory Requirements}


On GPUs, memory is primarily managed in 2 MB page frames for data \cite{tunnels, nayak2021mis}. We find that page table pages (page directory and page table) are also allocated within contiguous 2 MB regions, which we call \textit{PT regions}. Hence, forcing the GPU to allocate a page table page at a specific location requires first exhausting the entire 2 MB PT region with page table pages. Given 16 B PTEs per 2 MB data page, this would require allocating roughly 256 GB of data, well beyond the capacity of even the latest NVIDIA Blackwell B200 GPUs (180 GB DRAM per GPU) \cite{nvidia_b200_spec}. 

While smaller page frame sizes exist for UVM (e.g., 64 KB or 4 KB), that would reduce this memory requirement, the driver provides no control to the application to select which page type should be allocated, and adopts proprietary algorithms to aggressively coalesce pages to larger 2 MB pages. 
To address this, we develop techniques to efficiently fill existing PT regions using smaller data page frames, and show \textit{how} to allocate new PT pages at arbitrary locations under practical memory constraints, in \cref{sec:pt_alloc_eff}.

\subsubsection*{Challenge 3: Allocations are Silent}

To massage PT regions to the desired locations, the attacker needs to know \textit{when} a new PT region allocation occurs, without any driver modifications. On CPUs, many OSes expose CPU page table memory consumption to users that can serve as an explicit signal for CPU PT allocations: for example, Linux reports leaf-level page table usage via the \texttt{PageTables} field in \texttt{/proc/meminfo}\cite{proc_meminfo}.
On GPUs however, the CUDA runtime or driver do not expose any such information about GPU page table memory consumption, 
preventing processes from knowing when page table allocations occur. 
To address this, we demonstrate a new timing side channel in UVM memory allocations, which can reveal when PT region allocations occur, in \cref{sec:pt_side-channel}.
\section{Techniques for Page Table Massaging}
\label{sec:pt_techniques}
To address the above challenges in page table massaging and understand \textit{where, how,} and \textit{when} page tables can be allocated, we develop the following techniques. 

\subsection{Inspecting Page Tables}
To understand GPU page table layouts, we instrument the NVIDIA GPU driver, like prior work~\cite{tunnels}, and dump the entire physical memory, and inspect page table locations for a GPU process. 
All our analysis is executed on the NVIDIA driver version \REVISION{580.95.05 (released a few months ago)}, but we confirm the same behavior on a wide range of major driver versions \REVISION{(545.23.08-580.95.05)}.

\begin{observation}
By default, GPU page table pages are allocated in \textbf{2~MB PT regions} away from data.
To allocate new PT regions at vulnerable locations, we need to fill existing PT regions, without impractical data memory usage.
\end{observation}


\subsection{Memory-Efficient PT Region Allocation}
\label{sec:pt_alloc_eff}

A naive approach to populate the 2~MB PT region with PTEs 
by allocating 2~MB data page frames, the default for \texttt{cudaMalloc}, is impractical.
As shown in \cref{tab:ratio}, with 2 MB page frames, each PD0 entry occupies 16 B, yielding a data-to-PTE memory ratio of $128\text{K}:1$, requiring around \textbf{256 GB} of data memory to fully populate a PT region.

While this overhead can be substantially reduced by using smaller page frames supported by UVM, the driver provides no control to applications over which page sizes are used on memory allocations. 
For instance, with 64 KB and 4 KB data page frames in UVM, each PT entry is 8~B, decreasing the data-to-PTE ratio to $8\text{K}:1$ and $512:1$, respectively. This reduces the required data memory to  \textbf{16~GB} and \textbf{1~GB}. 
Therefore, to populate PT regions efficiently, we develop techniques to coerce the driver into using 64 KB and 4 KB data pages for UVM allocations, without coalescing them into 2 MB pages. 

\begin{table}[htbp]
    \centering
    \caption{User memory required to fill a 2MB large page with PTEs, for different data page frame sizes.}
    \begin{tabular}{Sc|ScScSc}
    \toprule
    \textbf{ Data Page Frame Size} $\rightarrow$ & \textbf{2MB} & \textbf{64KB} & \textbf{4KB} \\
    \midrule
    PTE Size        & 16B  & 8B & 8B \\
    Ratio of Data to PTE Bytes & 128K : 1 & 8K : 1 & 512:1 \\
    \midrule
    \makecell{User Memory for 2MB PTEs} & 256GB & 16GB & 1GB \\    
    \hline
    \end{tabular}
    \vspace{0.2em}
    \label{tab:ratio}
\end{table}



\begin{figure}[ht]
\centering
\includegraphics[width=3.3in,height=\paperheight,keepaspectratio]{"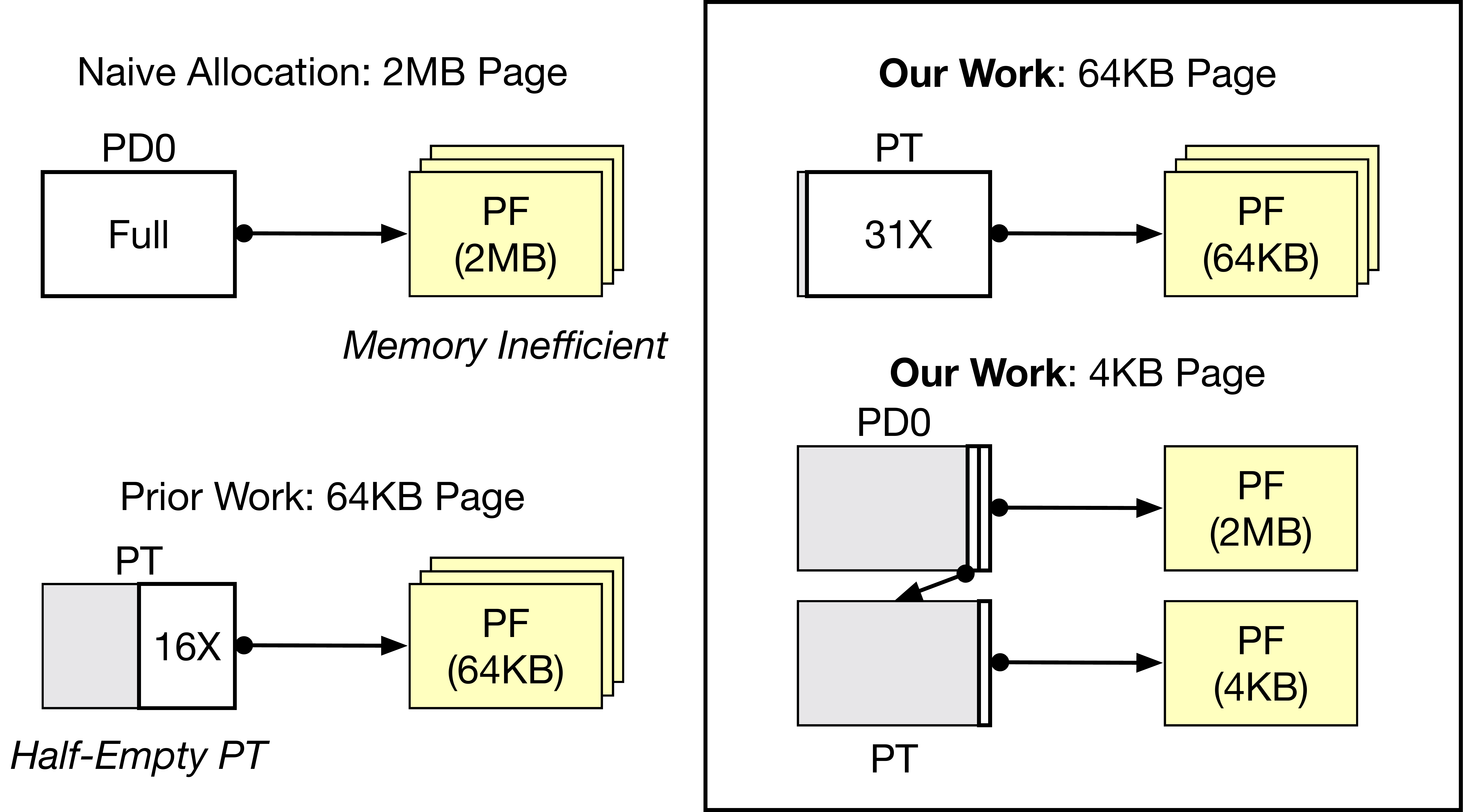"}
\caption{Page Table Filling Techniques. Prior techniques  are memory-inefficient or leave PTs half-empty. We uncover two techniques that address both deficiencies.} 
\label{fig:technique_efficient}
\end{figure}

\smallskip
\noindent
\textbf{Constraints.} 
While populating PT regions, we must satisfy two opposing goals: (1) \textit{memory-efficient PTE filling} for initial PT regions, minimizing the data memory needed, and (2) \textit{high-density PTE filling} for the PT region mapped to the vulnerable Rowhammer location, maximizing the chance that a bit-flip targets a valid PTE. 
However, no prior work has demonstrated how to allocate 4KB GPU data pages with UVM, the most memory-efficient option. 
Moreover, prior work~\cite{tunnels} shows that UVM allocated 64KB pages are automatically \textit{merged} into a 2MB page once more than 16 out of 32 of their PD0 entries are used, resulting in PT regions being half-empty.
Thus, naive filling of PT regions either waste memory (2MB frames) or yield sparsely populated page tables (64KB frames), as shown in  \cref{fig:technique_efficient} (left). We address both these limitations next.

\begin{figure}[ht]
\centering
\includegraphics[width=3.1in,height=\paperheight,keepaspectratio]{"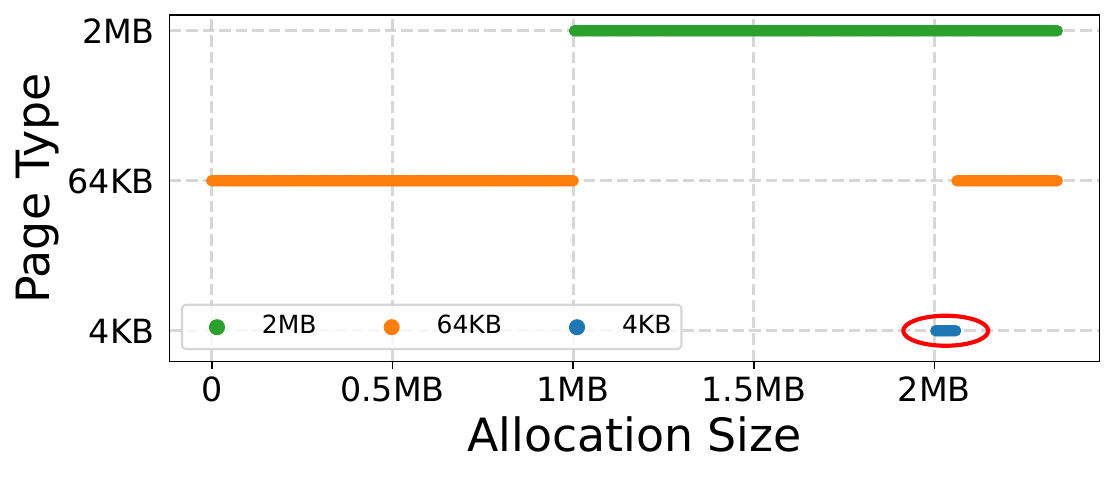"}
\caption{
Page types used as UVM allocation size vary. From $2\text{MB} + 4\text{KB}$ to 2MB + 64KB, 4KB pages are used.}
\label{fig:technique_page_type}
\end{figure}

\smallskip
\noindent
\textbf{(1) Memory-Efficient PTE Filling.} 
First, we develop techniques to fill the PT region with PTEs using 4KB page frames. 
We start by systematically analyzing the page-frame sizes used by the driver
for UVM allocations.
\cref{fig:technique_page_type} shows the page-frame sizes as UVM allocation size increases. We observe that for allocation sizes till 1MB, 64KB page frames are used, and then for 1MB to 2MB allocations, 2MB page frames are used. 
However, allocating $2\text{MB} + 4\text{KB}$ memory, results in both a 2MB and a separate 4KB page, as illustrated in \cref{fig:technique_efficient} (bottom right). 
Moreover, successive allocations of $2\text{MB} + 4\text{KB}$, result in new 4KB PT pages with a single valid PTE, corresponding to the 4KB page frame.
Thus, we achieve a ratio of data ($2\text{MB} + 4\text{KB}$) to PTE ($4\text{KB}$) bytes of $513:1$, allowing us to efficiently fill the 2MB PT region using  1GB of UVM allocated memory.\footnote{For the memory-efficient PTE filling procedure, for each $\text{2MB} + 4\text{KB}$ allocation, the 2MB portion can also be moved to the CPU side (where memory may be more abundant). This can reduce the GPU memory needed for this procedure to 2MB, suitable even for memory-constrained GPUs.} 

\smallskip
\noindent
\textbf{(2) Ensuring High Page Table Density.} 
4KB page frames are useful for filling initial PT regions, but not for filling PT regions at vulnerable memory locations.
This is because we see that one can only sparsely fill 16 out of 512 PTEs per PT page using 4KB page frames, before they start to get merged by the driver to 64KB pages. 
Although 64KB frames provide a denser PT population, prior work~\cite{tunnels} shows that these can also fill only 16 of 32 entries in a PT page (Figure 4, bottom left), leaving it \textit{half-empty} and reducing the chance that a bit-flip affects a valid PTE.

To address this, we introduce a technique that nearly fills the entire 256B PT page. We observed that GPU evicts UVM memory to CPU in 64KB chunks. 
Hence, instead of directly allocating 64KB pages, we allocate a 2MB UVM page and then access a 64KB slice from the CPU, forcing that slice to be evicted to system memory.
This causes the 2MB UVM page to be splintered into $32 \times 64\text{KB}$ pages, with 31 remaining on the GPU and 1 on the CPU, resulting in a densely filled PT page with 31 valid PTEs out of 32.

With this technique, we densely fill PTEs in approximately $97\%$ of the 2MB PT region containing the vulnerable location with the bit-flip, ensuring successful PTE tampering, while consuming  16GB memory.
Other 2MB PT regions (e.g., the initial PT region) can remain sparsely filled, requiring 1GB memory per 2MB PT region.

\begin{observation}
Thus, the attacker can fill the GPU PT region, either \textbf{memory-efficiently} (for the initial 2MB PT region) or \textbf{densely, with PTEs} (near vulnerable memory). 
\end{observation}



\subsection{Influencing PT Region Placement}
\label{sec:pt_placement}

To tamper page tables with Rowhammer-based bit-flips, the attacker requires its data pages to be co-located with PT regions in neighboring rows.
To enable this, we first understand the allocation heuristics of PT regions.

\smallskip
\noindent
\textbf{Allocation Heuristics.}
By observing the memory dumps of GPU processes, we empirically observe the PT regions are allocated between 64-128MB away from data pages.
However, once we deplete the initial 2MB PT region, using the memory-efficient PT allocation techniques from \cref{sec:pt_alloc_eff}, we observe subsequent PT regions are allocated from the same memory pool as data pages, enabling co-location of data pages and PT pages in neighboring DRAM rows. 


\begin{figure}[ht]
\centering
\includegraphics[width=3.3in,height=\paperheight,keepaspectratio]{"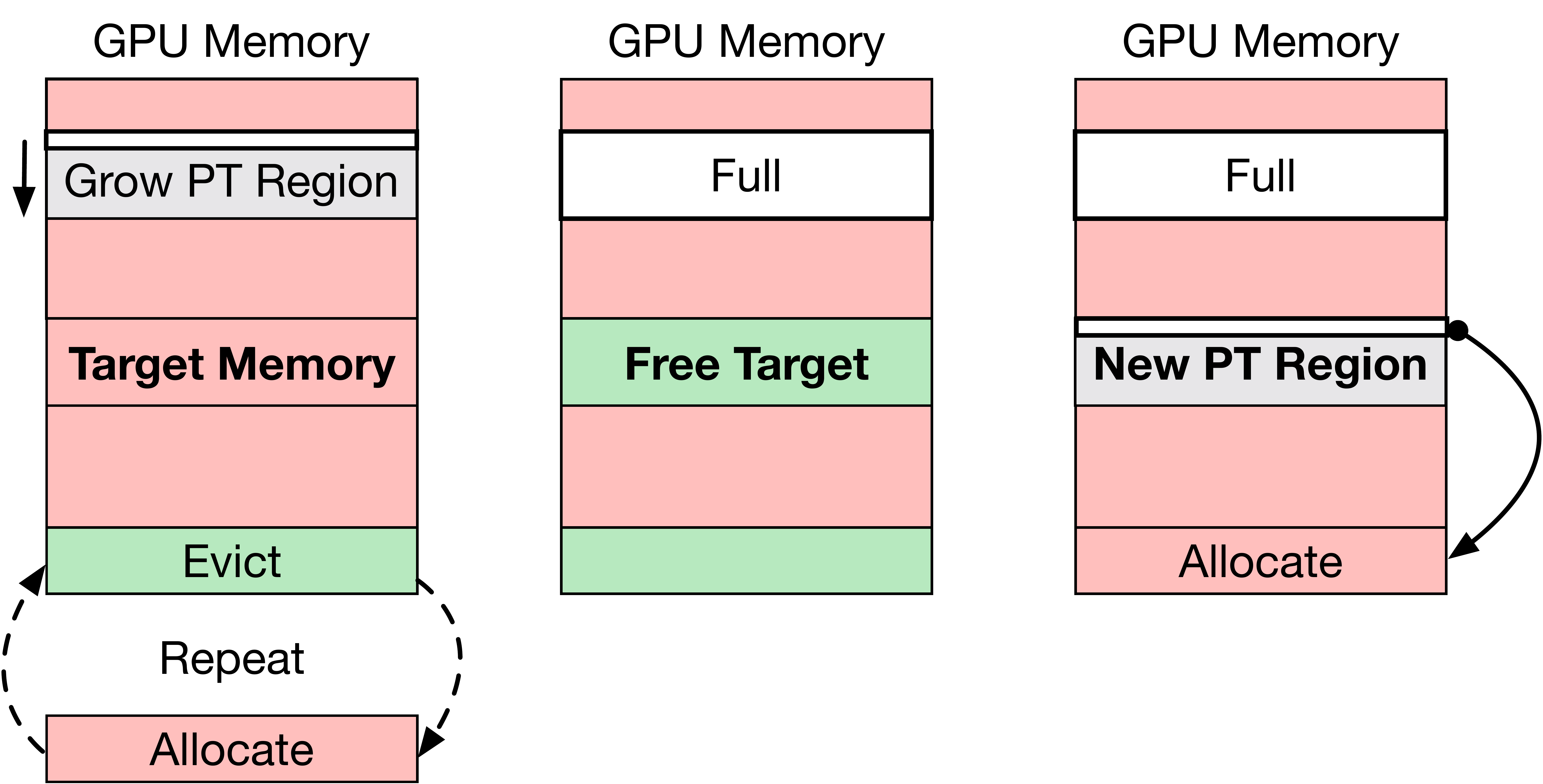"}
\caption{Workflow for Page Table Massaging.}
\label{fig:technique_massaging}
\end{figure}


\noindent
\textbf{Massaging PTEs to Vulnerable Locations.} 
Since regions are allocated from the same pool as user data pages, we use this to steer a PT region into a chosen physical 2 MB frame vulnerable to bit-flips. Our goal is to reach an allocator state that is one allocation away from creating a PT region, and only two page frames are free: one, the 2MB frame vulnerable to bit-flips, and another page frame for the data memory allocation, as shown in \cref{fig:technique_massaging}~(center). This accommodates both the new PT region to be allocated at the vulnerable location and the data memory allocation that triggers the PT creation.



\cref{fig:technique_massaging} illustrates our massaging technique. First, we exhaust device memory via UVM allocations.
We then repeat an evict-reallocate cycle, iteratively evicting a selected attacker data page from the GPU to the CPU and reallocating a new data page in its place, advancing the PT allocator to the point when it will create a new 2MB PT region. 
Before the next allocation, 
we evict the 2MB target data page (the page frame known to be vulnerable), by accessing it from the CPU, and evict another data page (2MB, 64KB or 4KB) elsewhere. 
The subsequent memory allocation forces the allocator to place the new 2MB PT region in the target slot, following the eviction order.
\begin{observation}
By \textbf{creating holes in the memory}, one at the vulnerable page frame and another elsewhere for a data frame, and \textbf{triggering a memory allocation creating a PT region}, the PT gets placed at the vulnerable location in memory.
\end{observation}
This PT region can now be filled densely with PTEs, as described in \cref{sec:pt_alloc_eff} for PTE tampering.

\begin{figure}[th]
\centering
\includegraphics[width=3.3in,height=\paperheight,keepaspectratio]{"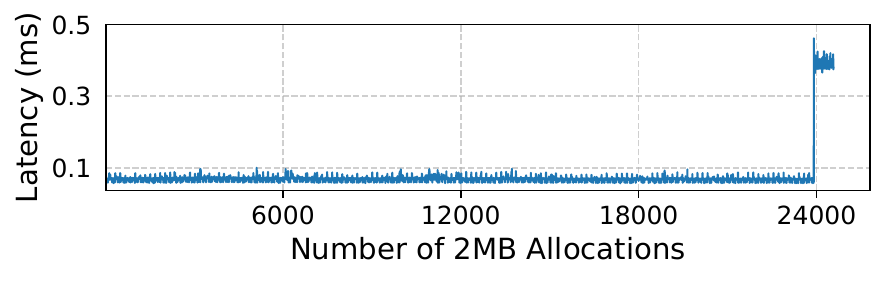"}
\caption{Spike in access latency when UVM memory allocation exceeds GPU DRAM capacity (48GB), due to page evictions from GPU to CPU.
}
\label{fig:technique_latency_full}
\end{figure}

\begin{figure}[th]
\centering
\includegraphics[width=3.3in,height=\paperheight,keepaspectratio]{"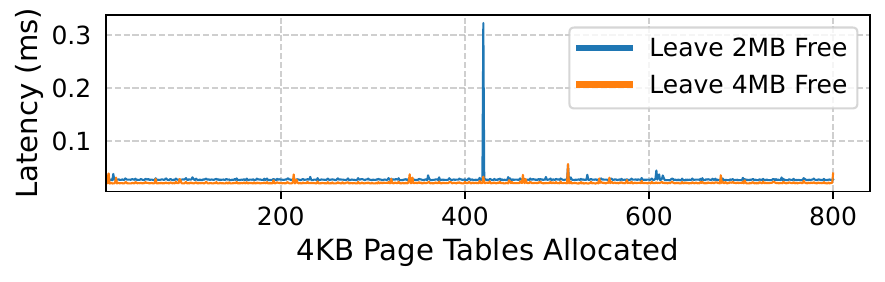"}
\caption{Spike in access latency when memory allocation causes 2MB PT region creation, if we leave only one page-frame free (e.g., 2MB free). No spike if free memory  can fit both data and PT region allocations (e.g., 4MB free).
}
\label{fig:technique_latency_comp}
\end{figure}

\begin{figure*}[!htb]
    \centering
\includegraphics[width=7in]{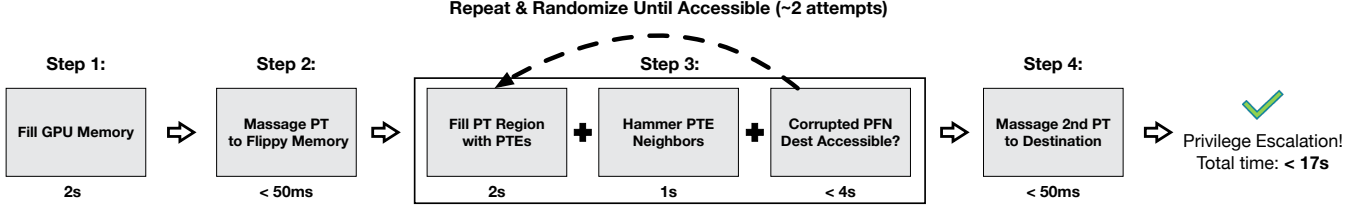}

    \caption{Workflow for the end-to-end attack on page tables to achieve GPU privilege escalation. \REVISION{Step 3 has a success rate of 44\% ($n=50$ attempts). The total exploit time takes 17s on average, including execution of Step 3 on average 2.3 times.}\leftrevisionbox{M1}}
    \label{fig:attack_flowchart}
\end{figure*}

\subsection{Page Table Page Allocation Side-Channel}
\label{sec:pt_side-channel}
For the attack, a critical requirement is that an unprivileged process knows exactly \textit{when} a new 2MB PT region is allocated. While a new PT region will be allocated every 1GB of memory allocation (as shown in \cref{sec:pt_alloc_eff}), the initial state of the PT region can be hard to predict, and vary based on the GPU process context and code size, making it hard to exactly know when new PT region allocations occur without instrumenting the driver. For this, we leverage a new timing side-channel emanating from UVM memory eviction. 

When the GPU memory is full and a new UVM memory page is to be allocated, a GPU page is evicted to the CPU. This eviction causes a high access latency for the first access, as shown in \cref{fig:technique_latency_full}, allowing us to precisely identify \textit{when} a UVM page eviction occurs from the GPU.
We leverage this page-eviction timing side-channel to infer when new 2MB PT regions are allocated. By keeping just one page-frame free on the GPU DRAM before a memory allocation,
which also requires a PT region allocation, it results in an eviction-caused spike in access latency, as shown in \cref{fig:technique_latency_comp}.
This allows the attacker to identify when the \textit{first} 2MB PT region allocation occurs.
Subsequent PT region allocations occur periodically after each 1GB memory allocation.

\begin{observation}
UVM page evictions introduce measurable timing side-channels. By \textbf{triggering page evictions on PT-region allocations}, an attacker can observe the timing variations to learn \textit{when} PT regions are created.
\end{observation}

\section{Attacking GPU Page Tables} \label{sec:attack}
In advance, the attacker profiles the memory for vulnerable locations susceptible to bit-flips  by executing Rowhammer campaigns similar to prior work~\cite{gpuhammer}.
Subsequently, using page table manipulation techniques from \cref{sec:pt_techniques}, we launch an end-to-end attack on GPU page tables using Rowhammer, to achieve privilege escalation.
\cref{fig:attack_flowchart} shows the steps (1-4) for the end-to-end attack, as described next.

\subsubsection*{Step 1: Fill GPU Memory}
The first step of the attack is to fill the entire GPU memory using UVM allocations, since the page-table massaging relies on a page-eviction side channel that requires all GPU memory to be occupied.
Unlike \texttt{cudaMalloc}, that fails when out of memory, UVM swaps memory to the CPU once the GPU DRAM is exhausted.
As the initial occupancy of the GPU DRAM is unknown, we detect when the memory is full, by measuring access latency after each 2 MB UVM allocation and checking for a latency spike that indicates a page eviction to CPU (\cref{fig:technique_latency_full}).
Once the attacker observes consecutive spikes, they know the GPU DRAM is fully allocated.



\subsubsection*{Step 2: Massage PT to Vulnerable Memory}\label{subsubsec:attack_step2}
Next, to enable PTE tampering, we force PTEs to be allocated in the physical page frames vulnerable to Rowhammer bit-flips. 
This consists of two stages: (1) Filling the first PT region and (2) Massaging the next PT region allocation to vulnerable frames. First, we fill the initial 2MB PT region with our memory-efficient PTE filling techniques, using $2\text{MB}+4\text{KB}$ pages, as discussed in \cref{sec:pt_alloc_eff}.
To monitor when the PT region is full, we iteratively free up a 2MB region of the GPU DRAM (by touching it from the CPU and swapping to the CPU), and then allocate 512 $\times$ 4KB page frames on the GPU (allocating $2\text{MB}+4\text{KB}$ UVM each time and accessing only the 4KB on the GPU side) as shown in  \cref{fig:technique_efficient}, and monitoring the access latency.
On a timing spike, as shown in \cref{fig:technique_latency_comp}, we know a new PT region has been allocated by evicting an existing page.\footnote{We observe the UVM evictions are in LRU order; so we preserve the vulnerable page frame in the GPU memory by re-accessing it periodically.} 
In this PT region, each 2MB + 4KB allocation allocates 32B in a PD0 entry and a 4KB PT page.
Thus, the number of $2\text{MB} + 4\text{KB}$ allocations ($A$) required to fill up the current 2MB PT region and create the next PT region is given by,
\begin{equation}\label{eq:1}
    2\,\text{MB} = 4\,\text{KB} \cdot (A + \lceil A/128 \rceil)
\end{equation}

\begin{figure}[ht]
\centering
\includegraphics[width=3.3in,height=\paperheight,keepaspectratio]{"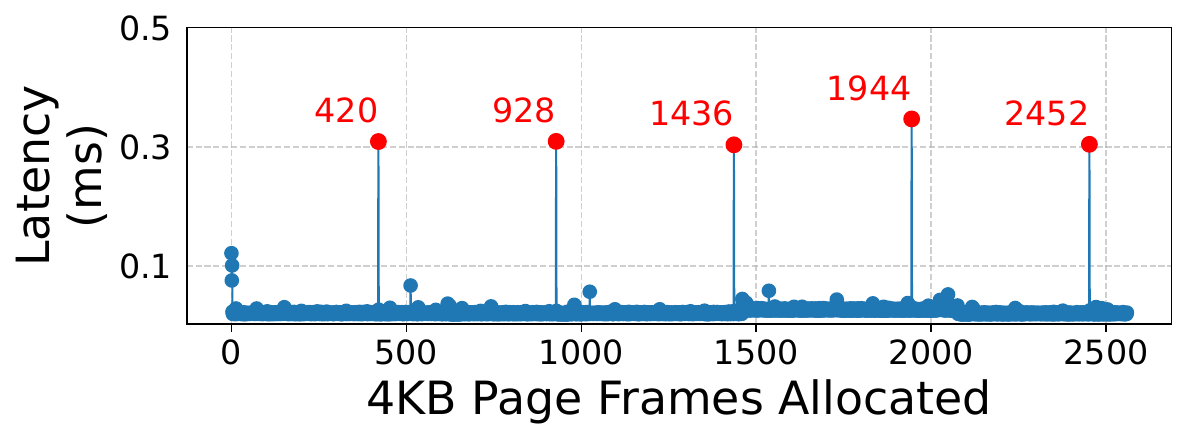"}
\caption{Access times after memory allocations of 4KB pages on the GPU. Spikes (red dots) occur when PT regions are allocated, every 508 data allocations, as per \cref{eq:1}.}
\label{fig:attack_4kb_latency}
\end{figure}

Based on \cref{eq:1}, we require 508 allocations in total for the next PT region to be allocated. 
We verify this empirically as shown in \cref{fig:attack_4kb_latency}, where we measure the access latency as the number of 4KB pages allocated on the GPU grows. 
The first spike corresponding to the first PT region allocation is after 420 allocations, and subsequent spikes appear every 508 allocations of 4KB frames.
Before the 508$^\text{th}$ allocation, we free up an additional 2MB frame by evicting the vulnerable page frame to CPU, and thus the next allocation is forced to map a new PT region to the target vulnerable 2MB frame. 


\subsubsection*{Step 3: Tamper PTEs and Check Corrupted PFN}

Next, we fill the 2MB PT region allocated to the vulnerable memory, densely with PTEs. As discussed in \cref{sec:pt_alloc_eff}, using 64KB data frames, we densely fill the PT region (31 out of 32 PTEs valid in each 256B PT page), ensuring the best chance of a bit-flip hitting a valid PTE.
Using techniques from GPUHammer~\cite{gpuhammer}, we hammer attacker-controlled data rows, neighboring the region where the PTEs are mapped, to flip the vulnerable bits and tamper PTEs.

\smallskip
\noindent
\textbf{Check Corrupted PFN Destination.}
After hammering, we check if this corrupted a PTE PFN, by sequentially reading through the attacker controlled data.
Initially, each data page has a sequential identifier written to it. 
On streaming through the data, if some data is out of order, we know the PFN of the VA was corrupted, and the data value tells us the correct VA for the accessed frame. If this occurs, we know the tampered PFN destination is attacker controlled ($x$).
Otherwise, if the data of the corrupted PFN destination is not recognizable, it may be an unassigned page, or a page of another user. 
In this case, we repeat Step-3; we randomly change the PFN values by evicting the corresponding data pages and re-allocating them in a different order, and repeating till the corrupted PFN destination is attacker controlled ($x$).
In practice, Step-3 achieves this in 2-3 attempts.


\subsubsection*{Step 4: Allocate PTEs at Corrupted PFN Destination}
Lastly, we evict the attacker-controlled page, $x$, that is at the corrupted PFN destination, and ensure another PT region is allocated into this frame.  
For this, we repeat the massaging process from Step 2. Once the PTEs are allocated to this physical frame, they can be accessed and modified using the VA of our previously corrupted PTE PFN. This completes our attack, as now we can arbitrarily control the PFN value for a given VA,\footnote{Each time we change the PTE PFN in DRAM, we need to evict its TLB entry, to ensure a page-table walk. We evict TLB entries by streaming through 2 GB of memory with 64KB pages which thrashes the TLB.}
through which we can access the entire GPU memory, thus achieving GPU-side privilege escalation.

\begin{observation}
Using the VA translated by the corrupted PTE, we access the second PTE and PFN2. By controlling PFN2, and using VA2 translated by PFN2, we can access the entire GPU memory, achieving GPU privilege escalation.
\end{observation}
\section{Exploitation Results}

To assess the potential for exploitation with \papername{} attacks, our evaluations answer the following questions:

\begin{enumerate}
    \item Can we leverage Rowhammer bit-flips to tamper PTEs and escalate privilege on a GPU? (\cref{subsec:exploit_pte_tampering_result})
    \item With our arbitrary GPU read primitives, can we leak sensitive data from GPU? (\cref{subsec:exploit_pqc},  \cref{sec:exploit_leak_weight})
    \item With our arbitrary GPU write primitives, can we tamper with GPU code, to enable stealthy accuracy degradation on ML models? (\cref{subsec:exploit_code_tamper})
    \item Can we use this to cross the device-host boundary and escalate privileges on the CPU? (\cref{sec:system-privilege-escalation})
\end{enumerate}

We answer these questions by studying \papername{} exploits on an NVIDIA RTX A6000 with 48 GB GDDR6 DRAM, known to be vulnerable to Rowhammer~\cite{gpuhammer}.

\begin{table}[h!]
\begin{center}
\begin{small}
\caption{\textcolor{\REVISIONTEXTCOLOR}{System Configuration}\rightrevisionbox{N3}}
\color{\REVISIONTEXTCOLOR}
\begin{tabular}{l|l}
\toprule
\textbf{Component} & \textbf{Specification} \\
\midrule
OS              & Ubuntu 22.04.5 LTS \\
CPU             & AMD Ryzen 5945WX \\
Compiler        & g++ 10.5.0 \\
GPU             & NVIDIA RTX A6000 \\
GPU Memory      & Samsung, 48 GB GDDR6 \\
NVIDIA Driver   & 580.95.05 \\
CUDA Toolkit    & 12.8 \\
\bottomrule
\end{tabular}
\label{table:system_config}
\end{small}
\end{center}
\end{table}

\subsection{GPU Privilege Escalation via PTE Tampering}\label{subsec:exploit_pte_tampering_result}
To achieve successful PTE tampering, the Rowhammer bit-flips in a PFN must change its destination to another 2MB page frame containing a PTE. This requires the PFN to change by at least 2MB, which requires the bit-flip to be in a range of 27 bits (PFN[34:8]) within the 48-bit PFN (PTE[42:16]), as per the PTE format shown in \cref{fig:pte}. 

\smallskip
\noindent
\REVISION{\textbf{Profiling Phase.} We use the code from  GPUHammer~\cite{gpuhammer}, to induce Rowhammer bit-flips on an RTX A6000 GPU. We generate the mappings of virtual addresses to banks and rows using row-buffer conflict side-channels~\cite{drama} (30~min/bank) and use synchronized multi-warp hammering patterns~\cite{gpuhammer,TRRespass,SMASH} to perform 24-sided hammering campaigns on 6 banks of the GDDR6 DRAM. We use the checkered data patterns, 0x55 and 0xaa, for the aggressor rows and their  inverse values for the victim rows. Our hammering campaigns take 3 hours per DRAM bank.}\rightrevisionbox{N3}

In the six DRAM banks we hammered (labeled $A - F$) on the RTX A6000, we observed 34 bit-flips in total, out of which 9 bit-flips are suitable for the privilege escalation with PTE corruption. \cref{tab:flip_info} shows the bit-flips that fit the criteria for privilege escalation with PTE tampering. The entire offline profiling phase takes \REVISION{$6$ banks  $\times$ $3.5$ hours $/$ $9$ flips $\approx3$~hours}\rightrevisionbox{N3}
per exploitable bit-flip.
\begin{table}[htbp]
\centering
\caption{Locations of Bit-Flips within 64-bit PTEs, suitable for GPU Privilege Escalation with PTE tampering}
\footnotesize{}
\begin{tabular}{c|cc|c|c}
\toprule
\multirow{2}{*}{\begin{tabular}[c]{@{}c@{}}Bit-flip\\ No.\end{tabular}} 
& \multirow{2}{*}{\begin{tabular}[c]{@{}c@{}}Byte\\Location\end{tabular}} 
& \multirow{2}{*}{\begin{tabular}[c]{@{}c@{}} Bit\\Location\end{tabular}} 
& \multirow{2}{*}{\begin{tabular}[c]{@{}c@{}}PTE Bit \\ $byte * 8 + bit$ \end{tabular}}
& \multirow{2}{*}{\begin{tabular}[c]{@{}c@{}}Jump\\ Distance\end{tabular}} \\
 &  & &  &  \\ \midrule 
    $A_1$ & 2 & 4 & 20  & 16 MB \\
    $B_1$ & 2 & 6 & 22 & 64 MB \\
    $C_1$ & 2 & 4 & 20 & 16 MB \\
    $C_2$ & 3 & 2 & 26 & 1 GB \\
    $D_1$ & 2 & 4 & 20 & 16 MB \\
    $E_1$ & 3 & 4 & 28 & 4 GB \\
    $F_1$ & 3 & 0 & 24 & 256 MB\\
    $F_2$ & 3 & 1 & 25 & 512 MB \\
    $F_3$ & 3 & 7 & 31 & 32 GB \\
    \bottomrule
\end{tabular}
\label{tab:flip_info}
\end{table}

\smallskip
\noindent
\REVISION{\textbf{Online Phase.}} Without loss of generality, we choose bit-flip, $A_1$, for our exploits, that produces the desired page-table corruption and analyze its exploitation end-to-end.
We execute Steps 1-4 in our attack workflow from \cref{sec:attack}, observing that we are required to repeat Step-3 two times on average, before we get a tampered PTE pointing to an attacker-controlled page, where we allocate a PT region.
Thus, by controlling its own PT, the attacker achieves arbitrary GPU R/W capabilities and device-side privilege escalation in less than 20 seconds.
This attack can run concurrently with other processes, in a temporally sliced manner, or spatially sliced with MPS. 


\subsection{Leaking Crypto Keys from GPU Memory} \label{subsec:exploit_pqc}
Using the arbitrary GPU read primitive from \papername{}, we show how an attacker process on the same system can recover secrets of another process resident on the GPU. As a case study, we leak private keys from NVIDIA’s post-quantum cryptography (PQC) library, cuPQC~\cite{cuPQC}, which accelerates PQC algorithms, and is used by 
libOQS~\cite{liboqs}.



\smallskip
\noindent
\textbf{Setting.} 
We target a desktop or workstation GPU, where the victim runs short-lived, GPU-accelerated key-exchange routines (MLKEM512) from cuPQC.
After each key-exchange routine, the victim copies its result back to CPU memory and the device memory is immediately freed.
The attacker has code-execution capabilities on this GPU and has gained arbitrary R/W privilege over GPU memory. 

\begin{figure}[thb]
\centering
\includegraphics[width=3.3in,height=\paperheight,keepaspectratio]{"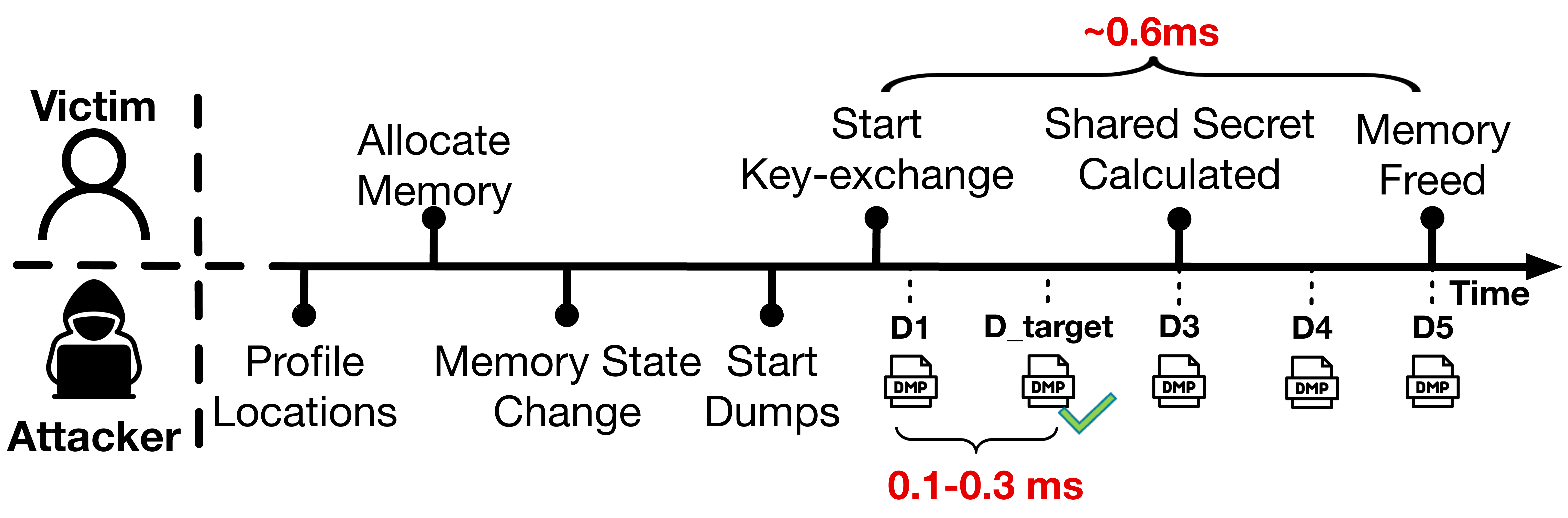"}
\caption{Timeline of attack on cuPQC}
\label{fig:exploit_cupqc_flow}
\end{figure}

\smallskip
\noindent
\textbf{Approach.} Dumping the entire device memory (takes more than 10 seconds) is too slow relative to the secret key residency on the GPU (less than a millisecond).
Instead, the attacker identifies a small candidate set of pages where the secret may be resident and performs rapid, targeted reads
using the approach in \cref{fig:exploit_cupqc_flow}, as we describe below.
\noindent
\textbf{1. Locating candidate pages.}
We observe that allocator placement of variables is largely reproducible across runs of the workload, and freed regions are zeroed by the runtime. Thus, to locate secret-resident pages, the attacker pre-fills the user pool with a non-zero pattern (e.g., \texttt{0xFF}). When the runtime later allocates, then zeroes freed regions, the zeroed pages stand out and form a short candidate list to search for the secrets (MLKEM512 keys occupy a few KB).


\noindent
\textbf{2. Detect Victim Allocations.} 
To detect when secret data becomes resident, the attacker monitors allocator state (e.g., via \texttt{cudaMemGetInfo}) to observe changes in free memory that correlate with victim allocations. Knowing when victim allocations occur, the attacker triggers targeted reads of our small set of candidate pages. 

\noindent
\textbf{3. Rapid targeted dumping.}
Targeted reads of a few megabytes are fast: on our RTX A6000, dumping 2MB takes about 0.1–0.3 ms, while an MLKEM512 key typically resides for around 0.6 ms. If the dump of the page containing the secret key overlaps this window, the secret is captured.


\smallskip
\smallskip
\noindent
\textbf{Results.} We run our exploit on the MLKEM512 example code in cuPQC.
To locate candidate victim pages, we prefill and scan the entire GPU user memory ($\sim$47 GB), requiring less than 5 minutes. After the victim ran, we observed $\sim$100 zeroed candidate pages. We applied our attack workflow, shown in \cref{fig:exploit_cupqc_flow}, against these pages, across 1000 independent victim key-exchanges.
We observe full recovery of the private key in 4.4\% of the key-exchange runs (44 out of 1000). This shows that an attacker with \papername{} can reliably leak short-lived cryptographic secrets from GPUs. 

In \cref{sec:exploit_leak_weight}, we similarly show how \papername{} can leak sensitive model weights from a long-running large language model (LLM) inference server.

\subsection{Stealthy ML Degradation via Code Tampering} \label{subsec:exploit_code_tamper}
Using the arbitrary-write primitive from \papername{}, we demonstrate GPU code tampering. Prior work showed that although bit-flips in ML model weights can degrade accuracy~\cite{gpuhammer,TBD,bitflipattack,Deephammer}, integrity checks can easily detect such corruption~\cite{NeuroPots}. Meanwhile, on CPUs, corrupting a single branch in BLAS code has been shown to stealthily degrade accuracy universally across all models~\cite{OneBitFlipMatters}. In this exploit, we show that stealthy code tampering is also possible on GPUs: tampering a single branch in proprietary cuBLAS on GPUs can produce hard-to-detect accuracy degradation universally across ML models. As GPU-resident code is not observable by users, such GPU code corruption is stealthier than weight corruption or even CPU code corruption.


\smallskip
\noindent
\textbf{Setting.} 
We assume the victim runs a long-lived inference process on the GPU. The attacker knows the victim’s PyTorch/cuBLAS versions and can identify targets in the code offline. The attacker seeks to induce \emph{silent} degradation universally across models by corrupting a \textit{single} branch in cuBLAS code under the following constraints: (1) the model must not crash, (2) outputs remain syntactically valid (e.g., a valid class), and (3) minimal latency impact (within 10\%).


\smallskip
\noindent
\textbf{Identifying Critical GPU Code.} 
We find that the GPU kernel code is resident in global memory in dedicated 2MB pages, and often ends with recognizable patterns like  \texttt{NOP} sleds, shown in \cref{lst:nop_sled}. We locate the cuBLAS code in the offline phase by taking the difference between code dumps before and after the cuBLAS library runs.

\smallskip
\begin{lstlisting}[language=C,  xleftmargin=1em,  xrightmargin=0.5em,  caption={\texttt{NOP} sled at the end of a GPU kernel code}, label={lst:nop_sled}]
EXIT ; //Exit current GPU thread
BRA 0x230; //Trap to prevent further execution.
NOP;NOP;NOP; ..........
\end{lstlisting}

\noindent
\textbf{Corruption Strategy.} Prior corruption used bit-flips to alter branch instruction directions.
In GPUs, branch instructions (BRA) as shown in \cref{lst:branch} are all unconditional, with the direction (taken or not) controlled by the predicate register (P0). With our arbitrary R/W primitive, we can prevent a branch being taken by simply \textit{replacing} \texttt{BRA} with a \texttt{NOP}. 

\smallskip
\begin{lstlisting}[language=C,  xleftmargin=1em,  xrightmargin=0.5em, caption={\texttt{BRA} instruction and predication in GPU}, label={lst:branch}]
ISETP.GE.AND P0, PT, R0, R2, PT; //R0 >= R2
PLOP3.LUT P0, PT, P0, PT, PT, 0x8, 0x0 ; //Mask
@P0 BRA 0x1d0 ; //Branch if P0 is TRUE (R0 < R2)
BRA 0xd0 ; //Branch Otherwise (R0 >= R2)
\end{lstlisting}

\begin{figure}[thp]
    \centering
    \includegraphics[width=\linewidth]{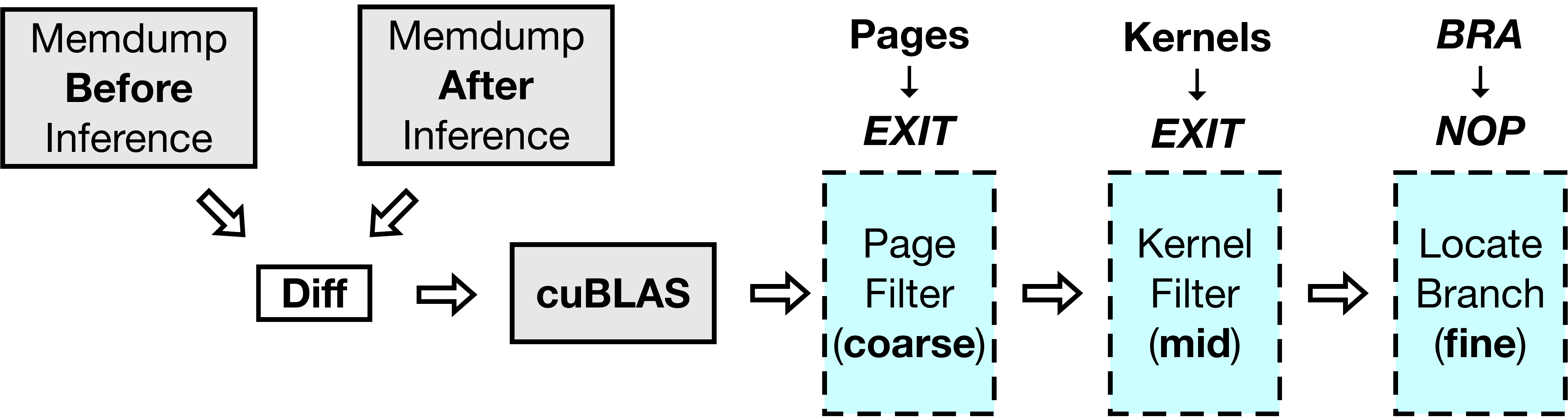}
    \caption{Filtering pipeline to locate vulnerable branches with coarse (page), mid (kernel), and fine (branch) reduction.}
    \label{fig:exploit_ml_filter}
\end{figure}

\noindent
\textbf{Vulnerable Branch Discovery.} We analyzed PyTorch with five image-classification models and found that the cuBLAS code contains over 5000 branch instructions. Furthermore, we observe that tampering with randomly selected \texttt{BRA} instructions is ineffective: even 
after 1000 attempts (24 hours) of tampering a randomly selected branch, we cannot find a vulnerable branch that universally reduces inference accuracy. 
To find such vulnerable branches efficiently, we develop a 3-stage filtering pipeline, as shown in \cref{fig:exploit_ml_filter}:


\begin{itemize}
  \item \textbf{Page filter.} We overwrite instructions in each 2\,MB code page with the \texttt{EXIT} opcode from \cref{lst:nop_sled}. Because \texttt{EXIT} immediately terminates a kernel, this allows us to quickly filter out code pages without significant impact on model accuracy when removed.
  \item \textbf{Kernel filter.} In pages that survive the page filter, we isolate individual kernels by detecting \texttt{NOP} sleds placed at kernel boundaries. We then iteratively replace half of the remaining kernels with \texttt{EXIT}, at each step, to identify a subset of kernels that are critical to accuracy.
  \item \textbf{Branch Identification.} Within each candidate kernel, we iteratively replace half of the remaining branch instructions with \texttt{NOP}s in each step and observe the inference, converging quickly on the critical branch.
\end{itemize}
\noindent
\textbf{Evaluation Results.} We ran the offline phase with PyTorch’s pretrained AlexNet and benchmarked its accuracy on a subset of the ImageNet dataset \cite{ImageNet} like prior works~\cite{gpuhammer, TBD}. From the cuBLAS binary, we observed a total of 5436 candidate branches. Applying the filtering strategy in \cref{fig:exploit_ml_filter} removed 75\% of those branches at the page filter, and a further 98\% at the kernel filter, leaving only 2 surviving kernels with 25 branches. From this reduced set, we try tampering with each branch, and locate a single branch that corrupts the models the most, with a minimal impact to inference latency. Our offline search process completes in under 100 benchmark runs across all stages, taking less than 2 hours. As shown in \cref{tab:stealthy_ML_code}, across all ImageNet models, tampering a single branch instruction universally degrades the accuracy of all models from 57-80\% down to 0.1\%. As this attack causes $<$6.6\% change in runtime, and the GPU code tampering is undetectable to users without access to device-side code, 
this attack is much stealthier than prior attacks tampering ML model code on CPUs~\cite{OneBitFlipMatters}, or model parameters on CPUs or GPUs~\cite{gpuhammer, TBD, bitflipattack, Deephammer} that are both easily detectable. 

\begin{table}[h]
\centering
\caption{All evaluated models show substantial accuracy degradation, with latency delta within $\pm$10\% of the original.}
\label{tab:stealthy_ML_code}


\resizebox{3.3in}{!}{
\begin{tabular}{l|cc|c}
\toprule
\textbf{Network} & \textbf{Base Acc.} & \textbf{Degraded Acc.} & \textbf{$\Delta$Latency} \\ 
\midrule
\textbf{AlexNet} & 56.68\% & 0.12\% & +1.26\% \\
\textbf{VGG16} & 72.20\% & 0.10\% & -4.32\%\\
\textbf{ResNet50} & 80.28\% & 0.10\% &  -6.62\% \\
\textbf{DenseNet161} & 77.20\% & 0.10\%  & -2.92\%\\
\textbf{InceptionV3} & 69.90\% & 0.10\% & -6.64\%\\ 
\bottomrule
\end{tabular}
}
\end{table}


\subsection{CPU-Side Privilege Escalation}\label{sec:system-privilege-escalation}

Thus far, \papername{} has shown elevated privileges on the GPU-side with arbitrary read/write capabilities on GPU memory. 
Prior work has shown that compromised PCIe devices~\cite{thunderspy} 
can alter the host state via DMA and obtain escalated privileges on the host.
In this section, we demonstrate how a compromised discrete GPU, such as in \papername{}, can also tamper with driver memory on the CPU to escalate to root level privileges on the system.

\smallskip 
\noindent
\textbf{Setting.} 
We assume a workstation GPU with a single user, or a GPU in a containerized setting in the cloud, shared between containers of different users~\cite{GKE,AlibabaCloudGPU}.  
We assume the attacker has arbitrary R/W privilege on the GPU memory via \papername{}, which can write to GPU PTEs. As shown in \cref{fig:pte}, GPU PTEs have a 2-bit Aperture field, which indicates the memory domain the PTE refers to. When the Aperture is set to the system memory mode (\texttt{0b11}), the PTE refers to CPU memory, and the PFN is interpreted as a CPU address. 
Thus, by overwriting GPU PTEs (aperture and PFN), the attacker can point to and access CPU memory. 

\begin{figure}[htbp]
    \centering
    \begin{subfigure}[b]{0.44\columnwidth}
        \centering
        \includegraphics[width=\textwidth]{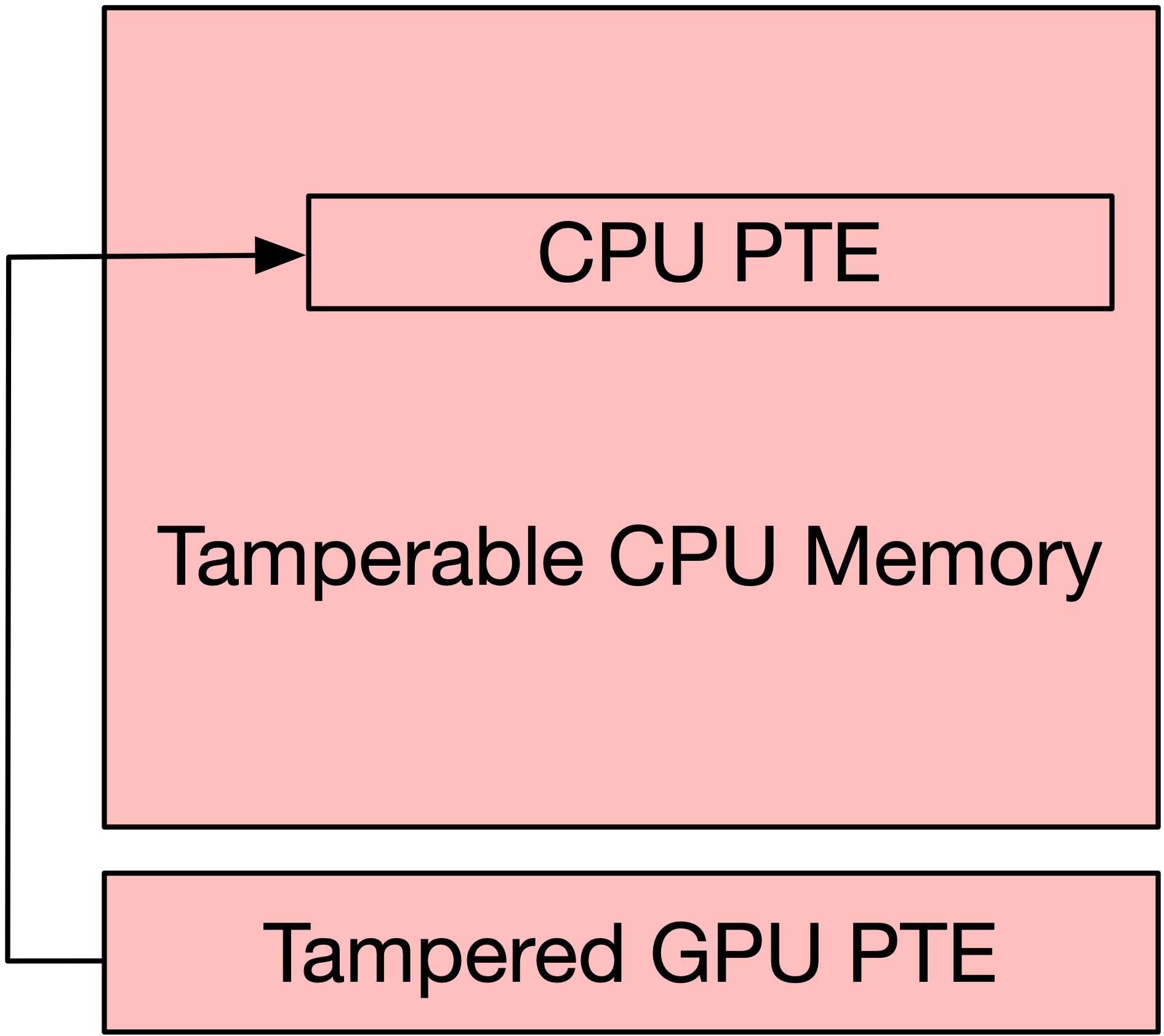}
        \caption{IOMMU off}
        \label{fig:cpu-privesc-non-iommu}
    \end{subfigure}
    \hfill
    \begin{subfigure}[b]{0.44\columnwidth}
        \centering
        \includegraphics[width=\textwidth]{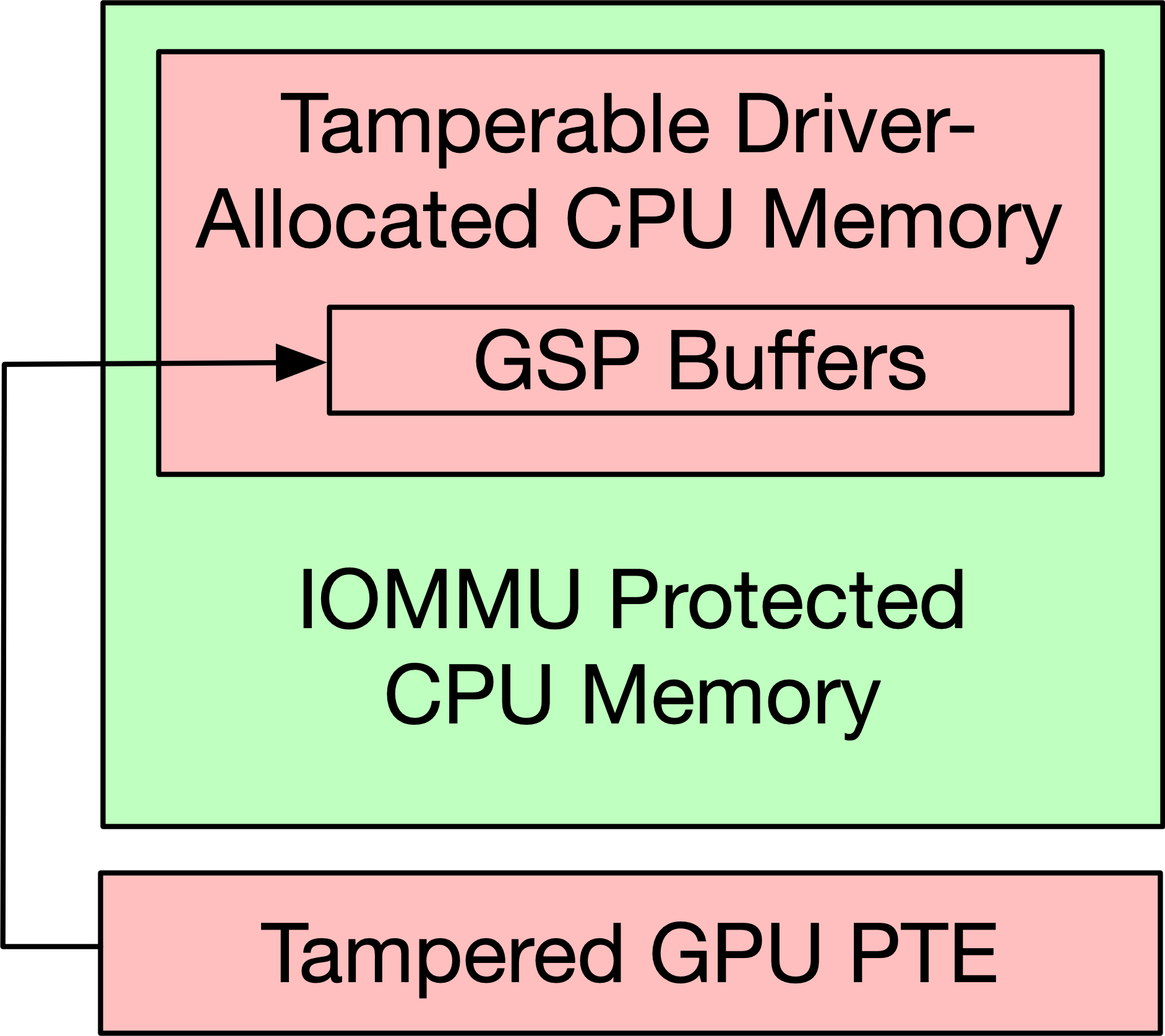}
        \caption{IOMMU on}
        \label{fig:cpu-privesc-iommu}
    \end{subfigure}
    \vspace{0.1in}
    \caption{CPU memory accessible by a compromised GPU. Without IOMMU, the entire CPU memory is accessible from GPU. With IOMMU enabled, the GPU can only DMA into driver memory on CPU assigned to the device.}
    \label{fig:tamperable-cpu-memory}
\end{figure}

We also assume the system enables the IOMMU, as recommended by all major vendors including NVIDIA and AMD~\cite{nvidia_iommu,amd_iommu, msft_iommu}.
IOMMU is a \emph{standard defense} against DMA-based attacks, which
prevents a malicious device from accessing arbitrary host memory.
Thus a compromised GPU is restricted to only the IOMMU-permitted driver memory on the host (the IOVA region mapped by the IOMMU), as shown in \cref{fig:tamperable-cpu-memory}. With IOMMU enabled, when a PTE has the Aperture bits set to system memory mode, the PFN corresponds to the IOVA. 
\rightrevisionbox{N2}
Finally, we assume that the attacker can recover kernel \REVISION{address information, such as} KASLR offsets through other side-channels~\cite{liu2023entrybleed,KASLRFormalDead,PrefetchSideChannel}.

\smallskip
\noindent
\textbf{Attack Overview.} 
The goal of the attack is to corrupt the kernel-mode GPU driver, running on the CPU, within IOMMU-permitted regions, to gain root-privileges.
\REVISION{The attack achieves this by corrupting trusted data-structures within IOMMU-permitted regions in the NVIDIA kernel-mode driver, that triggers memory-safety bugs in the driver, leading to arbitrary write primitives within the linux kernel.}

\REVISION{The driver allocates shared buffers in CPU memory for GPU communication via DMA, including RPC queues, fault buffers, and semaphores. These buffers are written exclusively by privileged GPU components (GSP and GMMU) and read by the kernel driver. We find that these GPU-produced values are systematically trusted without validation, flowing into security-sensitive operations as unchecked memcpy sizes, loop bounds, and array indices.} A compromised GPU can thus overwrite messages written by the GSP\REVISION{/GMMU} in these buffers, before the driver consumes them, creating 
\REVISION{memory safety issues} within the kernel, eventually leading to system-wide privilege escalation.

Through manual inspection, we find \emph{dozens of} \REVISION{sites} in the NVIDIA driver code where GPU-managed data structures
are consumed without sanitization. These buffers can be overwritten by a compromised GPU as the IOVAs for these structures are \REVISION{stable} even on system restarts.
\REVISION{Below, we describe the most interesting kernel OOB write, achieved by tampering the GSP message queue, which results in root-level privilege escalation.}
We cover the other vulnerabilities in the latter part of this section and in \cref{sec:other-oobs}.

\vspace{0.15in}
\noindent
\textbf{Overwriting GSP Message Queue.}
We tamper the GSP message queue structure in the GPU driver, which is used for device-to-host notifications, \REVISION{located at the offset \texttt{0x42000} from the base address of the IOVA region and searchable via the pattern of the queue's content.} The queue consists of 63 entries, each 4 KB in size; a message can span multiple consecutive entries directed by an \texttt{elemCount} member in the first entry. \REVISION{Once the GSP writes a message into the message queue, it will increment the \texttt{rxAvail} counter by one.} 
The driver \REVISION{checks if there are enough messages according to the value of \texttt{rxAvail} and \texttt{elemCount}}. If there are enough messages, the driver copies messages from the GSP queue into a kernel staging buffer (maximum 16 elements).
We observe an important metadata structure, \texttt{pMetaData}, is placed immediately after this staging buffer. It contains critical fields, like callback function pointers and other pointers used to access and modify data. GSP's input \texttt{elemCount}, is trusted and not sanitized before use. Hence, a malicious GPU can tamper this variable to any value larger than 16 causing OOB writes out of the staging buffer, which will overwrite members of \texttt{pMetaData}.

\newpage

\REVISION{By supplying a crafted \texttt{elemCount} in the $N$-th entry of the GSP message queue, an attacker can force the driver to copy beyond the staging buffer and overwrite \texttt{pMetaData} with attacker-controlled data. This requires passing the driver’s validation checks on the shared counter \texttt{rxAvail}.
To pass the driver’s validation checks, the attacker needs to inflate \texttt{rxAvail} to at least 17 to indicate sufficient entries are available. This can be achieved by flooding the queue with repeated \texttt{cudaDeviceGetAttribute} calls, causing the GSP to enqueue messages faster than the driver consumes them. We describe how we select the value of the queue index $N$ in \cref{sec:privesc-implementation}.}

\begin{figure}[!htp]
    \centering
    \includegraphics[width=0.9\linewidth]{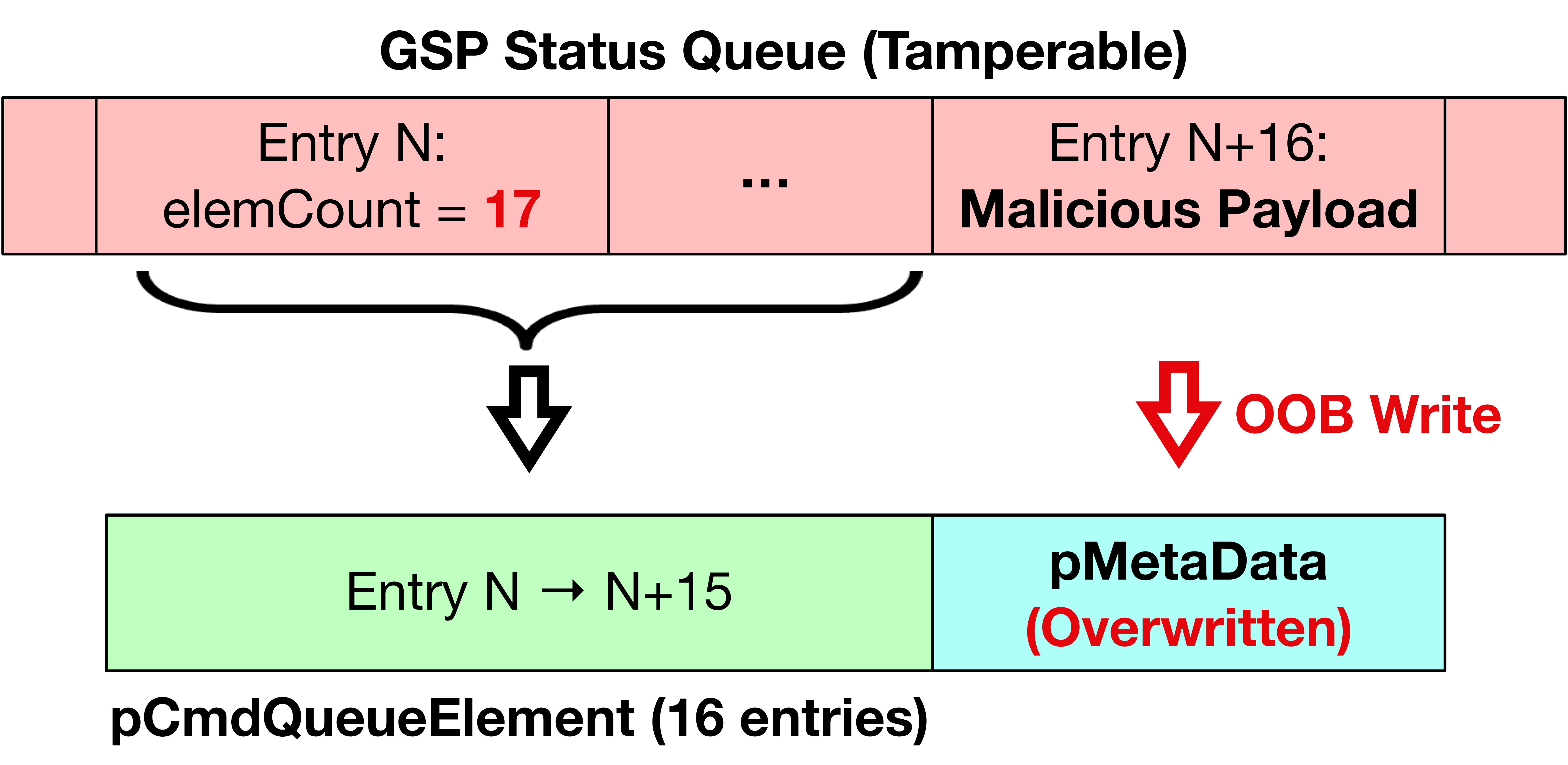}
    \caption{CPU-side privilege escalation. The compromised GPU injects \texttt{elemCount=17} and a malicious payload in the driver memory accessible via IOVA, that causes an OOB write in GPU inaccessible driver memory, that overwrites \texttt{pMetaData}. \texttt{pMetaData} contains data pointers used later, that when overwritten, result in arbitrary 4-byte writes in the kernel.}
    \label{fig:pmetadata-attack}
\end{figure}

Because the attacker controls all queue entries, they fully control the overwritten \texttt{pMetaData} (example payload in \cref{fig:pmetadata-attack}).
With this technique, the attacker has several gadgets in the GPU kernel driver to perform privileged operations, such as \textbf{arbitrary writes} and kernel-level execution via \textbf{arbitrary function calls} (\cref{sec:arbitrary-function-call}). We discuss one of the strongest primitives we found, a controlled \textit{4-byte write to an arbitrary kernel address}, and use it to escalate to root privilege on the host.

\noindent
\textbf{Arbitrary 4-byte Kernel Write \& Privilege Escalation.}
After overwriting the \texttt{pMetaData} structure by triggering an element-wise copy from the GSP queue into the staging buffer, the driver calls \texttt{msgqRxMarkConsumed}. 
In this function, the driver writes the 4-byte value of \texttt{rxReadPtr} into the address stored in \texttt{pReadOutgoing}.
By controlling the values of both members, the attacker can perform arbitrary 4-byte writes in kernel memory.
Using this primitive, the attacker can modify the target process's credential structure (struct \texttt{cred}) in the kernel and set the euid to 0, to get root-level privileges.  
\cref{sec:privesc-implementation} provides more details about the exploit, like the value of the malicious payload, and the steps to ensure kernel stability after the exploit.
Note that since the GPU message queue is corrupted in this exploit, a GPU reset is required to restore functionality. 
Crucially, the kernel remains in a stable state, enabling the attacker to execute arbitrary code with elevated privileges.

\begin{observation}
By exploiting DMA from the compromised GPU to the privileged GPU driver running on the CPU, we demonstrate the \textbf{first system-wide privilege escalation attacks} \textbf{using} \textbf{Rowhammer on discrete GPUs}, even with protections like IOMMU enabled.
\end{observation}

\vspace{0.1in}
\noindent
\REVISION{\textbf{Other OOBs by Tampering Fault Buffer.} The fault buffer in the NVIDIA driver is a potent target for tampering. It contains over a dozen GPU-produced fields read by the driver using the
\texttt{READ\_HWVALUE\_MW} macro, 
without runtime bound validation. We find that \texttt{UVM\_ASSERT} macros, used with some of these reads, are debug-only checks that are removed in release builds. 
The fault buffer fields such as \texttt{fault\_address}, \texttt{gpc\_id}, \texttt{client\_id}, and \texttt{fault\_type}, are later used in address calculations, indexing operations, and control-flow decisions.
If these values are maliciously overwritten, these can lead to OOB memory accesses. We describe some of these OOBs in \cref{sec:other-oobs}.}

\section{Discussion and Limitations}
\noindent \textbf{Applicability to Other GPUs.}
Our exploits are demonstrated on the NVIDIA RTX A6000 GPU, a widely used workstation and cloud GPU previously shown to be susceptible to bit flips~\cite{gpuhammer}. This is because the underlying Rowhammer vulnerability is GPU-specific, with attack patterns varying based on the memory vendor, in-DRAM defenses, and memory technology. 
Our PTE massaging based exploitation technique, however, applies to any NVIDIA GPU vulnerable to Rowhammer. NVIDIA’s security notice~\cite{NVIDIARowhammerNotice} indicates that GPUs across multiple generations (Ampere, Ada, Hopper, and Blackwell) may be affected by Rowhammer. 
\REVISION{We have verified that our CPU-side privilege escalation exploit, based on the driver vulnerabilities, remains effective across a wide range of driver versions (545.23.08–580.95.05), and is independent of the GPU model. We further validated it on two additional GPUs, a laptop RTX 3080-Laptop (Ampere) and a server L4 (Ada), using simulated bit flips in the absence of observable Rowhammer-induced bit flips on these GPUs.}\leftrevisionbox{N1}
Future studies that may discover Rowhammer bit-flips on other GPUs could also apply our exploits to those GPUs.

\vspace{0.1in}
\noindent \textbf{Impact of ECC.}
Our exploits assume ECC is disabled, consistent with real-world usage of many GDDR-based GPUs. 
We observe that cloud providers continue to offer A6000 instances with ECC disabled by default, as enabling ECC incurs a 6.25\% memory overhead and up to 10\% slowdown on GDDR-based GPUs~\cite{gpuhammer,IMTSullivan}. 
Enabling system-level ECC or using GPUs with on-die ECC in DRAM chips (e.g., GDDR7 or HBM3) may reduce the threat, but not fully prevent our exploits. 
Recent Rowhammer attacks~\cite{phoenix,eccfail} on CPU memories show that DRAM ECC mechanisms, which detect up to two-bit errors, fail to prevent Rowhammer attacks inducing three-bit errors or more, leaving systems exposed even with ECC enabled. 
Future work can evaluate such attacks under ECC-enabled configurations on GPUs.


\vspace{0.1in}
\noindent \textbf{UVM Requirement.}
Our exploits require 64KB and 4KB page frames for massaging PTEs, that are obtainable only through Unified Virtual Memory (UVM). 
UVM is supported on all NVIDIA GPUs since Pascal and is available in most multi-tenant environments, including Multi-Instance GPUs (MIG), Multi-Process Service (MPS), containerized deployments using Kubernetes~\cite{GKE,AlibabaCloudGPU}, and GPU confidential computing. 
Future works can explore these environments.
Our exploits do not apply when UVM is currently disabled, such as in vGPU configurations supporting live migration, or when multiple vGPUs are time-sliced on a single GPU~\cite{nvidia-uvm-vgpu}.
\cref{tab:applicability} lists the settings where our exploit is applicable.

\begin{table}[ht]
\centering
\caption{\REVISION{Applicability matrix of the attack.}\leftrevisionbox{M3}}
\label{tab:applicability}
\color{\REVISIONTEXTCOLOR}
\vspace*{1em}
\color{\REVISIONTEXTCOLOR}\begin{tabular}{c|cccccc}
\toprule
 & \textbf{ECC} & \textbf{UVM} & \textbf{IOMMU} & \textbf{MPS} & \textbf{MIG} & \textbf{CC} \\
\midrule
\textbf{Off} & Yes & Yes & Yes & Yes & Yes & Yes \\
\textbf{On} & No & No & Yes & Yes & Unknown & Unknown \\
\bottomrule
\end{tabular}
\end{table}


\section{Mitigations}
\noindent \textbf{Prevent PT Massaging.}
Since our exploits rely on precisely massaging PTEs to desired locations,
restricting fine-grained control over PT placement can prevent these attacks. 
For instance, the GPU driver can statically isolate page tables from data pages to avoid their co-location in memory, which is required by our Rowhammer attacks~\cite{Citadel,Siloz}. While attacks like PTHammer~\cite{PTHammer} are theoretically possible, they are harder on GPUs due to higher memory latencies. Similarly, randomizing PT region locations can further hinder targeted placement. Additionally, limiting the use of small pages (4KB or 64KB) increases PT massaging memory requirements by 16-256$\times$, making such attacks infeasible on most GPUs with under 100GB of RAM, albeit at the cost of up to 256$\times$ higher latency for access to UVM pages across the CPU-GPU boundary. Disabling UVM similarly hinders PT massaging, but limits application functionality~\cite{Gali2024,GH200-UVM}.

\vspace{0.1in}
\noindent \textbf{Input Sanitization in the GPU Driver.} 
To prevent a compromised GPU from escalating privileges on the CPU, the GPU driver could treat all GPU-originated data consumed by it as untrusted. By routing inputs from the GPU through a staging buffer and enforcing bounds and sanity checks before consumption, the driver can prevent memory-safety vulnerabilities in the kernel due to a compromised GPU and block CPU-side privilege escalation in our exploits.

\vspace{0.1in}
\noindent {\bf Adopt Principled Rowhammer Defenses.}
While enabling ECC can make attacks harder, adversaries can still craft successful exploits by injecting multi-bit-flips that surpass the ECC's detection capabilities~\cite{Eccploit,eccfail,phoenix}. 
Instead, integrity-based defenses that use Message Authentication Codes to detect PTE tampering~\cite{CSIRowhammer,PT-guard,SafeGuard} can fully stop our exploit.
GPUs could also adopt principled in-memory Rowhammer mitigations, such as secure in-DRAM trackers~\cite{park2020graphene, PRIDE, MINT, ProTRR}, or Per Row Activation Counting (PRAC) mechanisms~\cite{jedec_ddr5_prac,QPRAC,MOAT,Chronus,MoPAC,Panopticon}, to fully eliminate the underlying vulnerability.


\newpage
\section{Related Work}

\noindent \textbf{GPU Exploits.} 
Prior GPU exploits have leveraged residual GPU memory to leak rendered webpages and ML model outputs~\cite{steal_webpage, zhou2016vulnerablegpumemorymanagement, sorensen2024leftoverlocalslisteningllmresponses}. GPU side-channel attacks like GPU.zip~\cite{gpuzip}, Pixnapping~\cite{wang2025pixnapping}, and NVBleed~\cite{zhang2025nvbleed} leaked rendered pixels, while Spy in the GPU Box~\cite{SpyintheGPUBox} and LeakyDNN~\cite{LeakyDNN} leaked ML model parameters. Other works\cite{tunnels,not_so_refreshing} demonstrated covert-channels through GPU micro-architectural optimizations. GPU code is also vulnerable to buffer overflows that enable control-flow hijacking and ML model tampering \cite{CuCatch,studyofoverflow, GPUMemoryExploitationfunprofit}. 
Recent work~\cite{GpuCocoDemystified} showed that RPC queue metadata, used for CPU-GPU communication, can leak information in confidential computing settings. 
In contrast, \papername{} introduces the first privilege escalation attacks on GPUs via Rowhammer, enabling arbitrary read/write of  GPU memory and even elevated privileges on the CPU by tampering GPU driver data-structures, surpassing the capabilities of all prior GPU exploits.




\smallskip 
\noindent \textbf{CPU Privilege Escalation via Malicious Device.} 
Prior works~\cite{darksideofshader,adreno-gpu} have shown that \REVISION{vulnerabilities in integrated mobile GPU drivers (ARM Mali, Qualcomm Adreno) can allow unprivileged applications to corrupt kernel memory and achieve} privilege escalation on the CPU.
\REVISION{Thunderclap~\cite{markettos2019thunderclap} showed that malicious peripherals can exploit OS vulnerabilities via DMA accesses to achieve privilege escalation, even with IOMMU enabled, using a custom FPGA-based device attached to the victim system.}
Thunderspy~\cite{thunderspy} exploited the lack of firmware verification in Thunderbolt controllers: with physical access, attackers could reflash SPI firmware to bypass security checks, and achieve DMA-based privilege escalation.
\REVISION{DMARacer~\cite{johannesmeyer2025dmaracer} identified DMA-based race conditions in driver code, including TOCTOU bugs that can be exploited by malicious peripherals.}
In contrast \REVISION{to prior attacks that require driver bugs, custom hardware, or physical access}, we show that a \emph{discrete} GPU, compromised via Rowhammer from an unprivileged process, can tamper with  trusted GPU-writable buffers in the privileged GPU driver, achieving CPU-side privilege escalation.


\smallskip
\noindent \textbf{Other Rowhammer Exploits.}
Prior Rowhammer exploits targeted CPU-based memories. 
In DDR3, Flip Feng Shui (FFS)~\cite{FFS} used Rowhammer to compromise cryptographic code, while Drammer~\cite{Drammer} achieved privilege escalation on mobiles.
ThrowHammer~\cite{ThrowHammer} and Rowhammer.js~\cite{RowhammerJS}  demonstrated these over the network and JavaScript respectively.
TRRespass~\cite{TRRespass}, Blacksmith~\cite{Blacksmith} and SMASH~\cite{SMASH} bypassed in-DRAM TRR defenses in DDR4, while 
ECCploit~\cite{Eccploit}, ECC.fail~\cite{eccfail}, and Phoenix\cite{phoenix} defeated ECC in DDR3, DDR4, and DDR5.
Zenhammer~\cite{ZenHammer} showed AMD CPUs with DDR5 memories are vulnerable.
Grand Pwning Unit~\cite{GrandPwning} used integrated GPUs on mobile SoCs, sharing the CPU DRAM, to perform Rowhammer attacks on LPDDR3  DRAM.
In contrast, we are the first to demonstrate system-wide privilege escalation using Rowhammer on \textit{discrete} GPUs with GDDR6 memory.
Prior works also use Rowhammer to flip bits in ML models on CPU DRAM~\cite{TBD,Deephammer,OneBitFlipMatters,DNNFaultInjection,bitflipattack,tol2023dontknock,chen2021proflip} or GPU DRAM~\cite{gpuhammer}, degrading model accuracy.
Our exploits go beyond this by enabling \textit{arbitrary} leakage or tampering of ML model code and data on GPUs. 

\smallskip
\noindent \textbf{Rowhammer Mitigations.}
Most prior mitigations target CPUs.
Software-based approaches such as GuardION~\cite{van2018guardion}, CATT~\cite{brasser2017can}, RIP-RH~\cite{bock2019rip}, ZebRAM~\cite{konoth2018zebram}, Siloz~\cite{Siloz}, and Citadel~\cite{Citadel} isolate memory across security domains, while Copy-On-Flip~\cite{di2023copy} leverages ECC to detect and relocate attacked pages. These strategies could mitigate GPU-based attacks on page tables with NVIDIA driver support.
Hardware-based defenses, including Rowhammer trackers~\cite{park2020graphene,mithril,PRIDE,Hydra,ProTRR}, aggressor relocation~\cite{AQUA,Shadow,RRS,SRS}, delayed activations~\cite{Blockhammer}, refresh-generating activations~\cite{REGA_SP23}, and page table integrity mechanisms~\cite{PT-guard}, that were developed for CPUs, could also be adapted for GPUs; however this can introduce varying performance costs depending on GPU latency and bandwidth constraints.
\section{Conclusion}
We present \papername{}, the first Rowhammer-based privilege-escalation on NVIDIA GPUs with GDDR6 memories.
By developing efficient GPU page-table massaging techniques, we tamper with GPU page tables using Rowhammer, achieving arbitrary GPU R/W capabilities. We demonstrate practical attacks including data leakage, stealthy code tampering, and privilege escalation on the CPU even in the presence of protections like IOMMU. Our work underscores the need for effective Rowhammer mitigations in GPUs, 
and a re-evaluation of the trust placed in the GPU within kernel-mode GPU drivers.
\section*{Acknowledgments}
This research was supported by Natural Sciences and Engineering Research Council of Canada (NSERC) Discovery Grants under funding reference number RGPIN-2023-04796 and RGPIN-2018-05931, and an NSERC-CSE Research Communities Grant under funding reference number ALLRP-588144-23. David Lie is also supported by Tier~1 Canada Research Chair CRC-2019-00242. Any research, opinions, or positions expressed in this work are solely those of the authors and do not represent the official views of NSERC, the Communications Security Establishment Canada, or the Government of Canada.

\section*{Ethics Considerations}

We disclosed our attack to NVIDIA’s PSIRT on November 11, 2025, and subsequently also to Google, Microsoft, and AWS, and assisted with reproducing the issue and exploring mitigations.
No users or production systems were affected by our experiments that were conducted on local machines within isolated environments, using proof-of-concept cryptographic workloads and open-source models.
Upon completion of the coordinated disclosure process, we have publicly released all the code artifacts associated with this work 
to foster open science. Our artifacts are available at: \url{https://github.com/sith-lab/gpubreach}.


\section*{LLM Usage Considerations}
LLMs were used for editorial purposes in this manuscript,
and all outputs were inspected by the authors to ensure
accuracy and originality.

\bibliographystyle{IEEEtran}
\bibliography{refs}




\newpage
\begin{appendices}
\crefalias{section}{appendix}

\section{CPU Privilege Escalation Exploit Details}\label{sec:privesc-implementation}
The adversary uses two threads to repeatedly overwrite entry $N$ and $(N+16)\bmod 63$ of the \texttt{pStatusQueue}, racing the GSP so the driver twice consumes the injected payloads. The attacker must still satisfy the sanity checks in \cref{lst:sanity-checks}. To avoid an unstable kernel caused by the corrupted queue, the adversary should also terminate immediately after privilege escalation if possible.

\begin{lstlisting}[language=C,  xleftmargin=1em,  xrightmargin=0.5em, caption={Checks in \texttt{GspMsgQueueReceiveStatus}.}, label={lst:sanity-checks}]
if (_checkSum32(...) != 0)
// Retry if checksum fails.
if (pCmdQueueElement->seqNum != rxSeqNum)
// Retry if sequence number is wrong.
\end{lstlisting}


\noindent

\textbf{Scanning the correct sequence number and index $N$.} To satisfy the sequence-number check in \cref{lst:sanity-checks}, the exploit must predict the exact sequence value the driver will expect on its next dequeue. Since the driver’s \texttt{rxSeqNum} always advances to the sequence of the last successfully dequeued entry, we can recover this value by dumping the \texttt{pStatusQueue} via the tampered GPU PTE and locating the entry with the largest sequence number: let this entry be at index $i$ with sequence $s_{\max}$. The driver will next expect $s_{\max}+1$, which must be placed into entry $N = (i+1)\bmod 63$. Likewise, the payload bytes that will later be copied into \texttt{pMetaData} must be written into the last entry the driver uses for payload consumption, namely index $(i+17)\bmod 63$, corresponding to the original $(i+16)$ payload entry shifted forward by one to align with the upcoming sequence.

\begin{lstlisting}[language=C,  xleftmargin=1em,  xrightmargin=0.5em, caption={A subset of fields in \texttt{pMetaData}}, label={lst:pmetadata}]
// pMetaData is used as type msgqMetadata
NvU32 *pReadOutgoing;   NvU32 rxReadPtr;
msgqTxHeader rx {... NvU32 msgCount; }
msgqFcnNotifyRemote fcnNotify; void *fcnNotifyArg;
msgqFcnBackendRw fcnBackendRw;
msgqFcnCacheOp fcnFlush; msgqFcnBarrier fcnBarrier
\end{lstlisting}
\begin{lstlisting}[language=C,  xleftmargin=1em,  xrightmargin=0.5em,
 caption={4-byte write in \texttt{msgqRxMarkConsumed}.}, label={lst:msgqrxmarkconsumed}]
msgqMetadata *pQueue = (msgqMetadata*)handle;
pQueue->rxReadPtr += n; // n=17
if (pQueue->rxReadPtr >= pQueue->rx.msgCount)
    pQueue->rxReadPtr -= pQueue->rx.msgCount;
_backendWrite32(pQueue, pQueue->pReadOutgoing, &pQueue->rxReadPtr, ...);
\end{lstlisting}


\noindent
\textbf{Writing the 4-byte value with crafted payloads.} The definition of the structure pointed by \texttt{pMetaData} is in \cref{lst:pmetadata}. The attacker can control the 17th element in the queue to write the 4-byte value of \texttt{rxReadPtr} into the address stored in \texttt{pReadOutgoing} at Line 5 of \cref{lst:msgqrxmarkconsumed} by satisfying the following conditions: (1) set \texttt{pReadOutgoing} to point to a target location; (2) choose \texttt{rxReadPtr} and \texttt{rx.msgCount} so that the value actually written at the call-site equals a desired value after the driver’s arithmetic \REVISION{from Line 2 to Line 4}. For example, to write 0, one needs to set \texttt{rxReadPtr=0xfffffff9} and \texttt{rx.msgCount=10}; (3) set \texttt{fcnBackendRw} to 0 so the driver performs a write. 



   

\noindent
\textbf{Improving kernel stability.}
Because the injected data corrupts the message-queue state, leaving malformed entries risks unpredictable behavior when the driver later parses them. To prevent further instability after privilege escalation, we deliberately crash the driver by overwriting the \texttt{fcnFlush()} (the first callback after the write) with an invalid address. When invoked, this pointer triggers an immediate module crash, terminating the driver before other subsystems can consume the corrupted queue and reducing the chance of a full kernel crash post-escalation.

\begin{lstlisting}[language=C,  xleftmargin=1em,  xrightmargin=0.5em, caption={Callbacks in \texttt{msgqRxMarkConsumed} after the 4-byte write.}, label={lst:msgqrxmarkconsumed-callbacks}]
if (pQueue->fcnFlush)
    pQueue->fcnFlush(pQueue->pReadOutgoing, 4);
if (pQueue->fcnBarrier) pQueue->fcnBarrier();
if (pQueue->fcnNotify)
    pQueue->fcnNotify(1, pQueue->fcnNotifyArg);
\end{lstlisting}

\section{Arbitrary Function Calls}\label{sec:arbitrary-function-call}
In addition to the arbitrary 4-byte write, the attacker can overwrite callback function pointers in {\sloppy\texttt{pMetaData}}. Among them, the above \texttt{\seqsplit{msgqRxMarkConsumed}} invokes three callbacks after the 4-byte write if these pointers are configured (\cref{lst:msgqrxmarkconsumed-callbacks}), which enables the attacker to redirect execution under their control, for further exploitation. 

\section{Other Out-of-bound Accesses in the Driver}\label{sec:other-oobs}

The following are representative fault buffer fields that can be tampered with to induce OOB accesses.

\smallskip 

\noindent
\textbf{\texttt{gpc\_id}.} 
The fault buffer is shared between the GPU and CPU: the GPU writes fault entries via DMA, and the CPU driver reads them. When the driver parses the entry (\cref{lst:utlb-oob-use}), \texttt{gpc\_id} is used to calculate the \texttt{utlb\_id} field without validation: the \texttt{UVM\_ASSERT} is skipped in the release version. The computed \texttt{utlb\_id} is then used as an array index in \texttt{batch\_context->utlbs[]}, leading to OOB memory accesses when \texttt{gpc\_id} exceeds the number of GPCs on the GPU. We list the OOB accesses below.

\smallskip

\begin{lstlisting}[language=C,  xleftmargin=1em,  xrightmargin=0.5em, caption={OOB examples caused by \texttt{utlb\_id}.}, label={lst:utlb-oob-use}]
// In parse_fault_entry_common
buffer_entry->fault_source.gpc_id = READ_HWVALUE_MW(fault_entry...);
utlb_id = buffer_entry->fault_source.gpc_id...;
UVM_ASSERT(utlb_id < ...); // Ineffective
buffer_entry->fault_source.utlb_id = utlb_id;
// In fetch_fault_buffer_entries
current_tlb = &batch_context->utlbs[current_entry->fault_source.utlb_id];
++current_tlb->num_pending_faults; //OOB W
current_tlb->last_fault = current_entry; //OOB W
\end{lstlisting}

\noindent
\textbf{\texttt{fault\_address}.} Similarly, the \texttt{fault\_address} field is parsed from the fault buffer entry using the same mechanism. The \texttt{fault\_address} is reconstructed from \texttt{ADDR\_HI} and
  \texttt{ADDR\_LO} fields (\cref{lst:oob-fault-address}). When the driver processes the fault,
  \texttt{fault\_address} is used to calculate \texttt{page\_index}, causing OOB accesses.


% ...



\begin{lstlisting}[language=C,  xleftmargin=1em,  xrightmargin=0.5em, caption={An OOB example caused by \texttt{fault\_address}.}, label={lst:oob-fault-address}]
// In parse_fault_entry_common
NvU64 addr_hi, addr_lo;
addr_hi = READ_HWVALUE_MW(fault_entry, ...);
addr_lo = READ_HWVALUE_MW(fault_entry, ...);
buffer_entry->fault_address=(addr_lo+(addr_hi...;
// In service_fault_batch_ats_sub
page_index = (fault_address - sub_batch_base) / PAGE_SIZE;
uvm_page_mask_set(read_fault_mask, page_index);
\end{lstlisting}

\section{Leaking Model Weights from GPU DRAM}
\label{sec:exploit_leak_weight}
Our goal is to leak LLM weights using our arbitrary GPU-read primitive and identify the base model architecture. 
We assume the attacker and victim co-locate on the same GPU and victim runs a long-lived inference process. 

\smallskip
\noindent
\textbf{Method.} First, we build per-layer reference fingerprints (mean, standard deviation, fraction of zeros) for the base models by locating layer offsets (by adding unique paddings in test runs) and computing statistics over the dumps from known weights. For an unknown dump, we parse it  and calculate similar statistics, and identify the best match via weighted cosine-similarity score.

\begin{table}[htbp]
\centering
\caption{Cosine similarity score between LLM model fingerprints (base and fine-tuned models).
Values closer to 1 (highlighted in green) indicate highly similar models. 
}

\label{tab:llm_fingerprint}
\renewcommand{\arraystretch}{1.1}
\resizebox{\columnwidth}{!}{
\begin{tabular}{llcccc}
\toprule
& \textbf{Variant} & \textbf{Llama2} & \textbf{Llama3} & \textbf{Mistral} & \textbf{Gemma} \\
\midrule
\multirow{3}{*}{\begin{tabular}{@{}c@{}}
    Llama2\\(7B)
  \end{tabular}}         & Meta*       & \cellcolor{green!15}0.99 & -0.20 & -0.05 &  0.29 \\
                         & Nous        & \cellcolor{green!15}0.99 & -0.20 & -0.05 &  0.29 \\
                         & Meta, Chat  & \cellcolor{green!15}0.99 & -0.20 & -0.05 &  0.29 \\
\midrule
\multirow{3}{*}{\begin{tabular}{@{}c@{}}
    Llama3\\(8B)
  \end{tabular}}         & Meta*           &  0.06 & \cellcolor{green!15}1.00 &  0.07 & -0.02 \\
                         & Nous            &  0.06 & \cellcolor{green!15}1.00 &  0.07 & -0.02 \\
                         & Meta, Instruct  &  0.06 & \cellcolor{green!15}1.00 &  0.07 & -0.02 \\ 
\midrule
\multirow{3}{*}{\begin{tabular}{@{}c@{}}
    Mistral\\(7B)
  \end{tabular}}         & v1.0*           &  0.94 & -0.18 & \cellcolor{green!15}1.00 &  0.23 \\
                         & Instruct v1.0   &  0.94 & -0.18 & \cellcolor{green!15}1.00 &  0.25 \\
                         & OpenHermes 2.5  &  0.94 & -0.18 & \cellcolor{green!15}1.00 &  0.23 \\ 
\midrule
\multirow{2}{*}{\begin{tabular}{@{}c@{}}
    Gemma\\(7B)
  \end{tabular}}         & Google*          & -0.04 & -0.20 & -0.08 & \cellcolor{green!15}1.00 \\
                         & Google, Instruct & -0.03 & -0.16 & -0.04 & \cellcolor{green!15}0.97 \\
\bottomrule
\end{tabular}
}
\smallskip

\raggedright
\footnotesize
* Baseline models used to generate the reference fingerprint.


\end{table}

\noindent\textbf{Evaluation.} We test four 7B LLM families ($\sim$16 GB parameters). A complete dump takes $\sim$1.5 minutes while fingerprinting (statistics + comparison) takes 
$\sim$15 minutes. Results show that across families, we can reliably identify the model families (within-family similarity $>$0.99) from fine-tuned models, using LLM weights dumped.
Future works can explore attacks like recovering ML model functionality, and membership inference, from leaked LLM model weights.

\end{appendices}
\newpage 

\section{Meta-Review}

The following meta-review was prepared by the program committee for the 2026
IEEE Symposium on Security and Privacy (S\&P) as part of the review process as
detailed in the call for papers.

\subsection{Summary}
This paper explores how attackers can use Rowhammer bit flips for privilege escalation attacks on an NVIDIA GPU. To accomplish this, they developed techniques for massaging page tables into targeted locations in the GPU’s memory so that they could flip a bit inside of a page table on the GPU. Using this capability, they are able to write to memory on both the GPU and CPU, and demonstrate how to tamper with ML models, steal cryptographic keys, and gain escalated privileges on the CPU.
\subsection{Scientific Contributions}
\begin{itemize}
\item 5. Identifies an Impactful Vulnerability.
\item 6. Provides a Valuable Step Forward in an Established Field.
\end{itemize}

\subsection{Reasons for Acceptance}
\begin{enumerate}
\item Identifies an Impactful Vulnerability: This paper shows how using Rowhammer to induce bit flips on a GPU can result in GPU-side privilege escalation, which can then be used for CPU-side privilege escalation. Additionally, the discovery of the GSP message queue OOB vulnerability in the NVIDIA driver is a notable contribution to software security research.
\item Provides a Valuable Step Forward in an Established Field: Whereas prior work was limited to corrupting GPU applications, this paper shows that the implications of bit flips on GPU memory are more severe than previously understood, with consequences for the host CPU.
\end{enumerate}

\subsection{Noteworthy Concerns} 
\begin{enumerate} 
\item While the authors evaluated their privilege escalation exploit on multiple driver versions and GPUs using simulated bit flips, they only demonstrated bit flips and an end-to-end compromise on a single GPU. 
\item Their attacks for degrading ML models and extracting keys from PQcrypto are not particularly impactful, as the arbitrary read/write primitive they present subsumes both.
\end{enumerate}

\end{document}